\documentclass[
journal=jpclcd, 
manuscript=letter, 
layout=twocolumn
]
{achemso}

\usepackage{amsmath,amssymb}
\usepackage{graphicx}
\usepackage{float}
\usepackage{natbib}
\usepackage{color}
\usepackage{xcolor}
\usepackage[colorlinks=true,bookmarks=false,citecolor=blue]{hyperref}

\graphicspath{ {./Figures/} }

\newcommand{\br}{\mathbf{r} }
\newcommand{\bk}{\mathbf{k}  }
\newcommand{\bE}{\mathbf{E} }
\newcommand{\bD}{\mathbf{D} }
\newcommand{\bH}{\mathbf{H} }
\newcommand{\bB}{\mathbf{B} }

\newcommand{\bP}{\mathbf{P} }

\newcommand{\bp}{\mathbf{p} }

\newcommand{\ha}{\widehat{a} }

\newcommand{\bmu}{ \boldsymbol{\mu} }

\newcommand{\eps}{\varepsilon }

\renewcommand{\Im}{\frak{I}\mathrm{m} }

\renewcommand{\tilde} {\widetilde }
\renewcommand{\tilde}[1] {\widetilde{ #1 } }
\renewcommand{\widehat} {\hat }
\renewcommand{\widehat}[1] {\hat{ #1 } }

\newcommand{\dyad}[1] {\overset\leftrightarrow{#1}}

\newcommand{\elm}{electromagnetic }
\newcommand{\fp}{Fabry-P\'erot }

\title{Chiral Polaritonics: Analytic Solutions, Intuition and its Use}

\author{Christian Sch\"afer}
\email{christian.schaefer.physics@gmail.com}
\affiliation{MC2 Department, Chalmers University of Technology, Sweden}

\author{Denis G. Baranov}
\affiliation{Center for Photonics and 2D Materials, Moscow Institute of Physics and Technology, Dolgoprudny 141700, Russia}
\email{denis.baranov@phystech.edu}


\begin{document}

\begin{tocentry}
\centering\includegraphics[width=1\columnwidth]{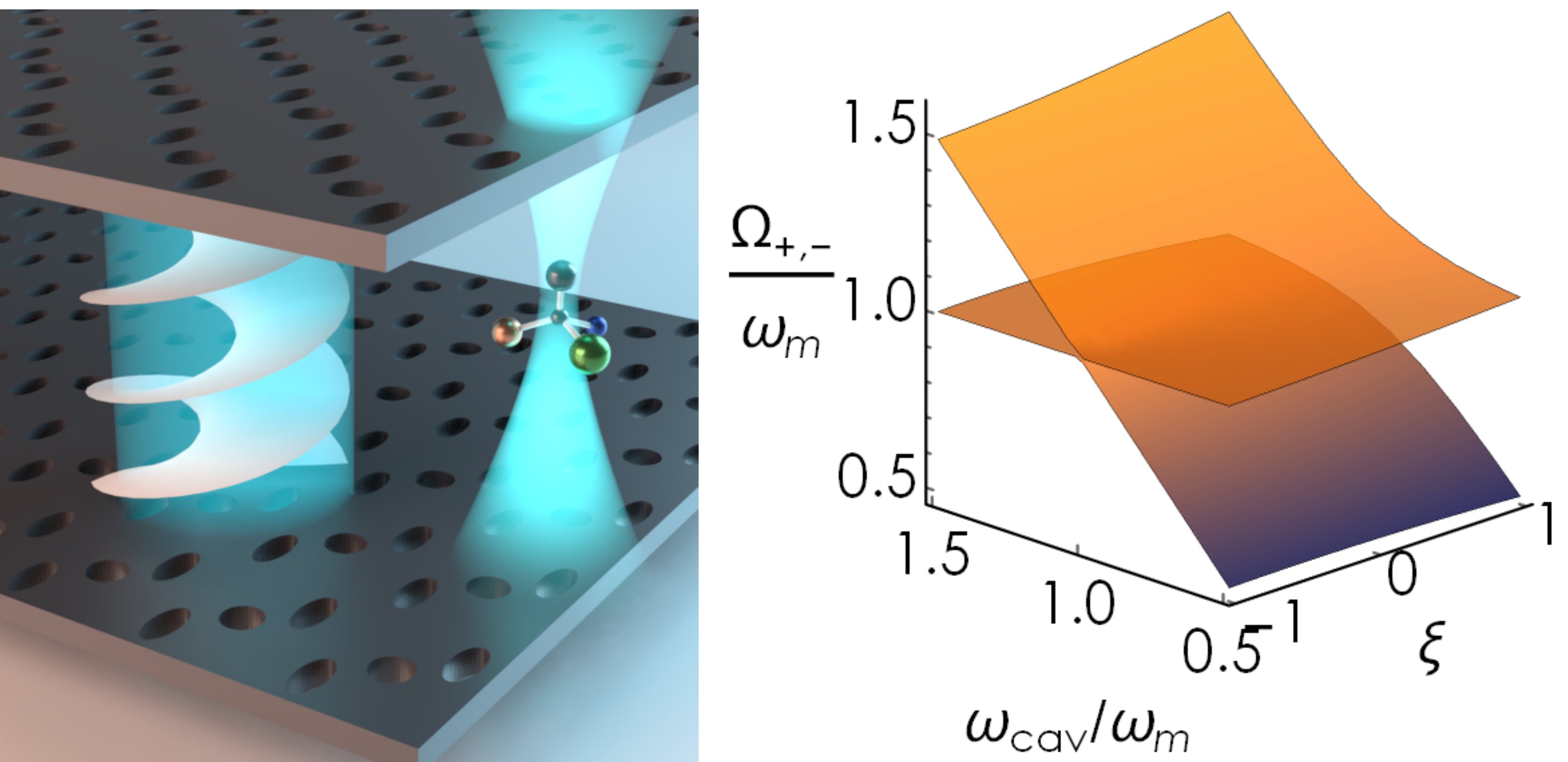}
\end{tocentry}

\begin{abstract}
Preferential selection of a given enantiomer over its chiral counterpart becomes increasingly relevant in the advent of the next era of medical drug design. In parallel, cavity quantum electrodynamics has grown into a solid framework to control energy transfer and chemical reactivity. In this work, we derive an analytical solution to a system of many chiral emitters interacting with a chiral cavity -- in analogy to the widely used Tavis-Cummings and Hopfield models of quantum optics. We are able to estimate the discriminating strength of chiral polaritonics, discuss possible future development directions, exciting applications such as elucidating homochirality, and deliver much needed intuition to foster the freshly flourishing field of chiral polaritonics.
\end{abstract}

\maketitle

Coupling between two harmonic oscillators, either of classical or quantum origin, leads to a hybridization and the creation of new quasi-particle states if the coupling strength exceeds all the decay and decoherence rates in the combined system. A common representative for such a system is the interaction between an energetically isolated electromagnetic mode and a set of quantum emitters such as molecules. The associated quasi-particle states are referred to as polaritons and possess mixed light and matter characteristics which opens a toolbox with enormous versatility \cite{garcia2021manipulating,simpkins2021mode,sidler2021perspective,forn2019ultrastrong,kockum2019ultrastrong,mennucci2019multiscale,luk2017multiscale,fregoni2021strong,bajoni2008polariton,chikkaraddy2016,wang2017coherent,baranov2019ultrastrong,hubener2021engineering,gubbin2020,doi:10.1021/acs.jpclett.1c01695}. Polaritons of various flavours have been used or proposed as path to enhanced charge and excitation transfer \cite{coles2014b,orgiu2015,zhong2017energy,schafer2019modification,du2018theory,hagenmuller2018,cohn2022}, modify chemical reactivity \cite{hutchison2012,thomas2019tilting,singh2022solvent,imperatore2021reproducibility,schafer2021shining,schafer2022emb,li2021collective,li2021cavity,galego2016,munkhbat2018suppression,groenhof2019tracking,kowalewski2016cavity,Vendrell2018,fabri2021probing,li2022semiclassical,fischer2022cavity}, and alter a systems state and response to external stimuli \cite{deng2003polariton,kena2010room,slootsky2014room,latini2021ferroelectric,flick2018light,schlawin2022cavity,schafer2021shortcut,lentrodt2020,doi:10.1063/5.0012754}, to name only a few.

So far, most experimental and theoretical efforts in this field have focused on coupling optical cavities with either linearly- or circularly-polarized electronic transitions of various quantum emitters. This is perfectly justified by the fact that in the visible and infrared ranges the interaction of light with electronic and vibrational transitions is dominated by the electric dipole term of the Hamiltonian. Nevertheless, there are examples of media that exhibit resonances with non-negligible magnetic transition dipole moment. One such practically relevant example is presented by the class of \emph{chiral} media \cite{Lindell2018,Barron2004,Condon1937}. 

A geometrical shape in three-dimensional space is called chiral if it cannot be aligned with its mirror image by a series of rotations and translations  \cite{Kelvin1894}.
The chirality occurs at various scales ranging from the shapes of galaxies down to drug and bio-molecules.
Especially the latter receives a steady stream of attention in the ongoing quest for new, safer and affordable ways to design chemical complexes and drugs \cite{weiskopf2002,CALCATERRA2018323}. 
It is then intuitively pivotal for its success to develop a solid understanding of relevant processes and a wide range of readily usable techniques that allow a separation or discrimination of the two enantiomers of a chiral structure.

While the recent years showed major progress in this field \cite{SCRIBA201656,WEINBERGER2000139}, sometimes referred to as chiral recognition, widely used chemical strategies such as (re)crystallization \cite{roberts1977basic} can be cumbersome and require often highly specified approaches for each individual compound.
Interaction of chiral matter with circularly polarized electromagnetic fields leads to the effect of circular dichroism, which underlies numerous methods for distinguishing molecular enantiomers \cite{Govorov2010}. However, those interactions are usually \textit{weak} and can be well understood without the need to consider a correlated motion between light and matter.
If and how \textit{strong} light-matter interaction can aid those challenging tasks remained largely unclear thus far.

While chiral polaritonics is still in its infancy, recent theoretical work is beginning to explore this question. Mauro et al. investigated the optical features of a single-handedness cavity loaded with a Pasteur medium using classical electromagnetism \cite{mauro2022chiral}. Riso et al. studied changes in the correlated ground state of single (or few) realistic molecules minimally coupled to an amplified chiral mode \cite{riso2022strong}.

\begin{figure}[h!]
\includegraphics[width=.5\textwidth]{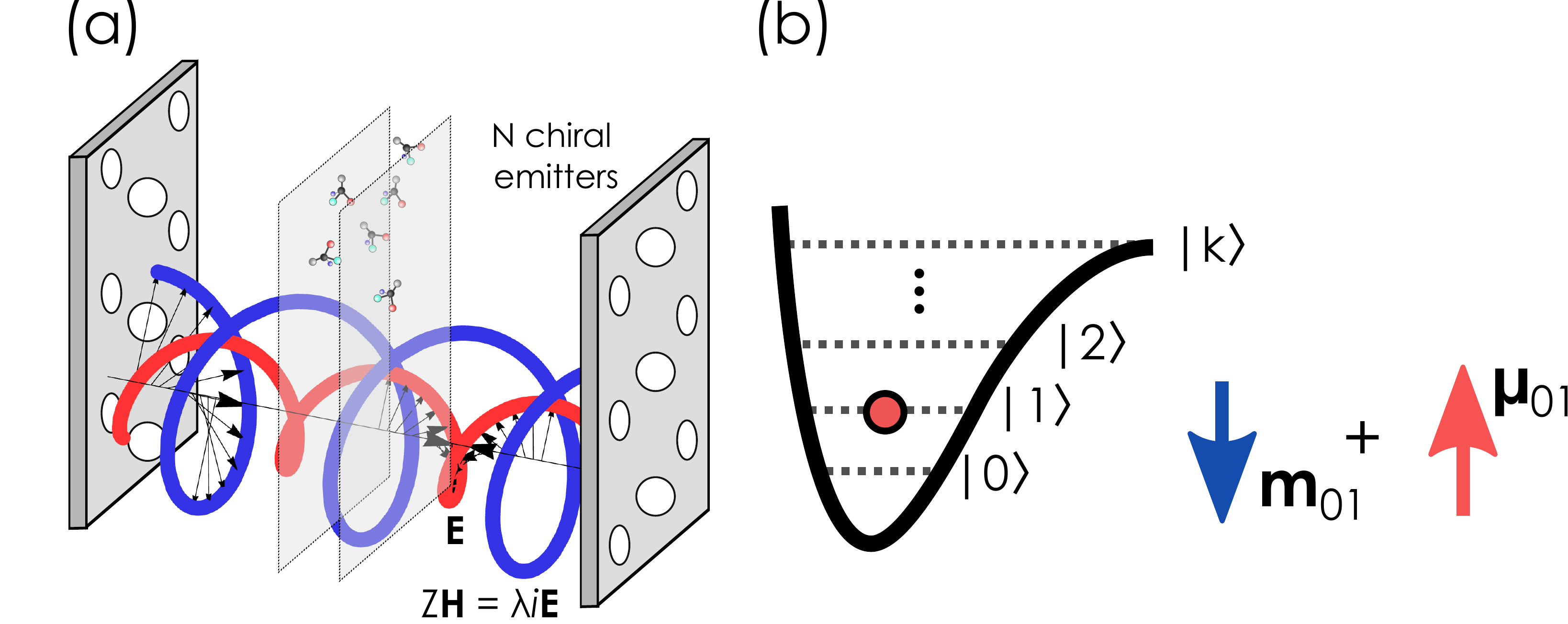}
\caption{(a) Illustration of the system under study. $N$ identical chiral (quantum) emitters interact with the \elm field of a chiral standing wave. A chiral standing wave is formed between two handedness-preserving metasurface mirrors. 
(b) A model of a single chiral emitter as generic $N$-level system. 
Chiral emitters, which represent molecules, biological structures or plasmonic meta-atoms, are modeled as simplified multi-level systems whose transitions are quantified by collinear electric and magnetic dipole moments.
}
\label{fig:concept}
\end{figure}

In this letter, we provide an analytic solution and much needed intuition that will be of good use for the future development of chiral polaritonics.
Starting from non-relativistic quantum electrodynamics (QED), we derive the quantized \elm fields supported by a single-handedness chiral cavity \cite{Voronin2022} and couple them to a large set of chiral emitters as illustrated in Fig.~\ref{fig:concept}. The resulting Hamiltonian can serve as starting point for any kind of \textit{ab initio} QED \cite{hubener2021engineering,schafer2021making,haugland2020coupled,haugland2020intermolecular,flick2017atoms,ruggenthaler2018quantum}. Here, we focus on a simplified model that allows for an analytic solution which illustrates the physical playground, potential, and forthcoming challenges of chiral polaritonics. In contrast to previous approaches \cite{WEINBERGER2000139,SCRIBA201656,mason2013optical,li2010theory,forbes2021orbital}, the strong coupling to a chiral cavity allows one to reach sizeable interaction strength even in the absence of any pumping field. The here derived model paves the way to invigorate an entirely new research domain.


We start our derivation from the non-relativistic and spin-less limit of QED in Coulomb gauge. Using the Power-Zienau-Wooley transformation and expanding the multi-polar light-matter interaction to second order introduces magnetic dipolar couplings and electric quadrupole terms \cite{babiker1974generalization,craig2004quantum,forbes2018,andrews2018perspective} according to
\begin{equation*}
    \hat{H} = \hat{H}_{M} + \hat{H}_{L} + \hat{H}_{LM},
\end{equation*}
where we differentiate between the $N_M$ electronic ($q_i=-e$) and nuclear ($q_i = eZ_i$) charges plus longitudinal Coulomb interaction among them constituting the molecule, and inter-molecular Coulomb interactions
\begin{align*}
    \hat{H}_{M} &= \sum_{n=1}^{N} \sum_{i=1}^{N_M} \frac{1}{2m_i} \hat{\bp}_{i,n}^2 + \sum_{n=1}^N \hat{V}^n_{\parallel}(\hat{\br}) + \sum_{n\neq n'}^{N} \hat{V}^{n,n'}_{\parallel}(\hat{\br}),\\
    \hat{H}_{L} &= \frac{1}{2} \int d^3 \br \left[ \frac{\hat{\bD}_\perp^2(\br)}{\eps_0} + \eps_0 c^2 \hat{\bB}^2(\br) \right].
\end{align*}
The interaction $H_{LM}$ up to magnetic order takes the form
\begin{align}
\label{eq:Hlm}
    \hat{H}_{LM} &= -\frac{1}{\eps_0} \sum_n \hat{\boldsymbol\mu}_n \cdot \hat{\bD}_\perp(\br_n) - \sum_n\hat{\textbf{m}}_n \cdot \hat{\bB}(\br_n)\notag\\ 
    &- \frac{1}{\eps_0} \sum_n \sum_{a,b\in \{x,y,z\}} \hat{Q}_{ab,n} \nabla_{a,n} \hat{D}_{\perp,b}(\br_n)\\
    &+ \frac{1}{8m} \sum_{n,i} (q_{n,i}\hat{\br}_{n,i} \times \hat{\bB}(\br_n))^2 + \frac{1}{2\eps_0} \int d^3 \br \hat{\bP}_\perp^2,    \notag
\end{align}
with the canonical particle momentum $\bp_i = m_i \dot{\br}_i - \frac{q_i}{2} \br_i \times \bB(\br)$, the total transverse polarization $\hat{\bP}_\perp = \sum_{n} \hat{\bP}_{\perp,n}$, the electric dipole moment $\hat{\boldsymbol\mu} = \sum_i q_i \hat{\br}_i$ and quadrupole $Q_{ab}$, as well as the magnetic dipole $\hat{\textbf{m}} = \frac{1}{2} \sum_i \frac{q_i}{m_i} \hat{\br}_i \times \hat{\bp}_i$. All positions are defined relative to each individual molecular center of mass.
Importantly, the multi-polar form introduces the displacement field $\hat{\bD}_\perp = \eps_0 \hat{\bE} + \hat{\bP}$ as the canonical momentum to the vector potential.
The last term in Eq.~\eqref{eq:Hlm} can be written in the tensorial form $ \sum_n \hat{\chi}^{m}_{n,ij} \hat{B}_{i}(\br_n)\hat{B}_{j}(\br_n) $ with $\hat{\chi}^{m}_{n,ij} = \sum_a q_a^2/8m_a (\hat{r}_{n,a}^{(i)}\hat{r}_{n,a}^{(i)} \delta_{ij} - \hat{r}_{n,a}^{(i)}\hat{r}_{n,a}^{(j)})$.

Let us briefly comment on the relevance of the appearing self-interaction contributions. The magnetic $ \sum_n \hat{\chi}^{m}_{n,ij} \hat{B}_{i}(\br_n)\hat{B}_{j}(\br_n) $ and electric $\frac{1}{2\eps_0} \int d^3\br \hat{\bP}_\perp^2$ self-polarization terms ensure gauge invariance and guarantee the stability of the correlated system \cite{schafer2019relevance,craig2004quantum}. A common simplification is to assume the molecules as well separated. In this dilute limit and when \textit{all} photonic modes are considered, the inter-molecular Coulomb interactions cancel perturbatively with the inter-molecular contributions arising from $\frac{1}{2\varepsilon_0}\int d^3\br \sum_{n,n'} \hat{\bP}_{\perp,n} \hat{\bP}_{\perp,n'}$ such that only retarded intermolecular interactions via the photonic fields remain \cite{power1982quantum,craig2004quantum}. However, when the number of photonic modes is truncated, as commonly done for polaritonic systems, this intuitive result does no longer hold. If the intermolecular contributions are neglected nevertheless, one arrives at the widely used Dicke model that falsely predicts a transition into a superradiant phase, while the associated model in the Coulomb gauge does not exhibit such a transition \cite{viehmann2011superradiant}. Under which conditions a phase-transition for more realistic systems could appear and what characterizes such a transition is still a matter of active debate \cite{latini2021ferroelectric,ashida2020,lenk2022,de2018cavity,nataf2019,andolina2020}. 
We provide a derivation in analogy to the common Dicke model in the SI and focus in the following on the development of a non-perturbative chiral model.

The electromagnetic fields follow the generic mode expansion
\begin{align*}
    \hat{\bD}_\perp(\br) &= i \sum_{\bk,\lambda} \sqrt{\frac{\hbar c k \eps_0}{2 V}} \big( \epsilon_{\bk\lambda} e^{i\bk \cdot \br} \ha _{\bk\lambda} - \epsilon_{\bk\lambda}^* e^{-i\bk \cdot \br} \ha ^\dagger_{\bk\lambda} \big),\\
    \hat{\bB}(\br) &= i \frac{1}{c} \sum_{\bk,\lambda} \sqrt{\frac{\hbar c k }{2 \eps_0 V}} \big( \beta_{\bk\lambda} e^{i\bk \cdot \br} \ha _{\bk\lambda} - \beta_{\bk\lambda}^* e^{-i\bk \cdot \br} \ha ^\dagger_{\bk\lambda} \big).
\end{align*}
where $V$ is the cavity mode volume, $\epsilon_{\bk, \lambda}$ and $\beta_{\bk, \lambda}$ are the electric and magnetic field unit polarization vectors; $\bk$ labels wave vectors, and $\lambda$ labels polarization states. 

Although Maxwell's equations in free space admit solutions in the form of chiral photons, in this case both handednesses coexist at the same time. Contrary to one's intuition, illuminating an ordinary \fp cavity with a circularly polarized light does not address this problem \cite{Baranov2020CD,hubener2021engineering}.
However, using cleverly designed  asymmetric "single-handedness" cavities \cite{Semnani2020,Voronin2022}, it is possible to engineer pure chiral \elm fields with only one handedness.
In a right-handed (left-handed) monochromatic wave the magnetic field is $\pi/2$ behind (ahead) the electric field everywhere in space $Z \bH(\br;\omega) = - i \lambda \bE(\br;\omega)$, where $Z=\sqrt{\mu_0/\varepsilon_0}$ is the impedance of free space, and $\lambda$ is the eigenvalue of the helicity operator \cite{Fernandez-Corbaton2016}, which takes values $+1$ and $-1$ for LH and RH fields, respectively. Only a subset of modes will adhere to the conditions that are imposed by the boundary conditions of the chiral cavity, it should be noted that the \elm fields are not zero at the mirror-surfaces.

A planar optical cavity, such as the one described in Ref. \cite{Voronin2022}, supports a continuous spectrum of resonant states that can be labels by their in-plane momenta $\bk_{\parallel}$. Cavity fields maintain their single-handedness quality for a substantial range of in-plane wave vectors (incident angles) \cite{Voronin2022}.
For simplicity, we will illustrate only coupling of a single standing wave with $\bk_{\parallel} = 0$ to the chiral emitters, and refer the interested reader to the SI for a generalized discussion.
The chiral standing wave is the superposition of two counter-propagating circularly polarized plane waves of the same handedness.  Assuming the axis of the cavity to be pointed along the $z$ direction and considering a vertical standing wave with $\bk = \pm k \textbf{e}_z$, the displacement field $ \hat{\bD}^{\lambda}_\perp(\br) = \sum_{\bk, \lambda} \hat{\bD}^{\lambda}_{\bk,\perp}(\br)$ of a LH/RH standing wave is 
$ \hat{\bD}^{\lambda}_\perp(\br) = (\hat{\bD}^{\lambda}_{+k,\perp}(\br) + \hat{\bD}^{\lambda}_{-k,\perp}(\br) )/\sqrt{2} $ with $\hat{a}_{+k,\lambda} = \hat{a}_{-k,\lambda}$ and
 $ \epsilon_{\pm k,\lambda}^{} = \frac{1}{\sqrt{2}}(1, \pm i \lambda, 0)^T$.

With these simplifications, the displacement field of a chiral standing waves takes the form
\begin{align*}
    \hat{\bD}^{\lambda}_\perp(\br) = - \sqrt{\frac{\eps_0}{V}} \tilde{\eps}^{\lambda}_{k}(z) \hat{p}_{k,\lambda},
\end{align*}
where $\tilde{\eps}^{\lambda}_{k}(z) =  \left(\cos(k z), - \lambda \sin(k z), 0 \right)^T$ is the z-dependent polarization vector of the chiral standing wave, and the canonical coordinates are $\hat{p}_{k,\lambda} = -i\sqrt{\hbar c k/2} ( \ha _{k,\lambda} - \ha ^\dagger_{k,\lambda}) $ and $\hat{q}_{k,\lambda} = \sqrt{\hbar/2ck} ( \ha _{k,\lambda} + \ha ^\dagger_{k,\lambda} )$. 
Notice that the left- and right-handed polarization vectors are orthogonal only in the spatially averaged sense.
Recalling the relation between the electric and magnetic field of a chiral field, $\beta_{\bk,\lambda} = -i \lambda \epsilon_{\bk,\lambda}$, we obtain the magnetic field of a chiral standing wave as $\hat{\bB}_{\lambda}(\br) = \sqrt{k^2/\eps_0 V} \lambda \tilde{\eps}^{\lambda}_{k}(z) \hat{q}_{k,\lambda}$.

The standing chiral fields satisfy Maxwell's equation and contribute with $ck=\omega_k$ a photonic energy of 
$\hat{H}_{L} = ( \hat{p}_{k}^2 + \omega_{k}^2 \hat{q}_{k}^2 )/2 =  \hbar \omega_{k} (\ha ^\dagger_{k}\ha _{k} + \frac{1}{2})$
for a given handedness. The extraordinary consequence is now that the standing field in an empty cavity will feature chiral quantum-fluctuations, i.e., for each Fock-state $\vert n \rangle$ the optical chirality density $C_n(\br,\omega) = \frac{\eps_0 \omega}{2} \Im \langle n \vert \hat{\bE} \cdot \hat{\bB}^{*} \vert n \rangle$ reduces to $C_n(\br,\omega) = \lambda \hbar \omega_k k / 4 V$.
Also a dark chiral cavity will influence the ground and excited states of matter located within it. In the following, we will introduce a series of simplifications and derive an analytic solution to the combined system of many chiral molecules coupled to the chiral cavity.


Let us here briefly describe the derivation of the analytical solution and their underlying models, a detailed version can be found in the SI.
For negligible intermolecular Coulombic interactions, i.e., $\hat{V}^{n,n'}_\parallel \approx 0 $, the molecular component to the Hamiltonian becomes diagonal $\hat{H}_{M} = \sum_{n=1}^{N}\sum_{k=1}^{\infty} E_{k}^{n} \vert k_n \rangle \langle k_n \vert $ in the many-body eigenstates $\vert k \rangle$. Expanding all transition elements in this eigenbasis would, in principle, allow for a numerical solutions of the (ultra-)strongly coupled system. The interested reader might refer to the area of \textit{ab initio} QED \cite{schafer2018insights}. Here, we focus on simplified models that provide analytical solutions.

The self-magnetization term mediated via $\hat{\chi}^m_{n,ij}$ will be assumed to be purely parametric $\hat{\chi}^m_{n,ij} \approx \chi^m_{n,ij}$ as it obstructs otherwise the Hopfield diagonalization scheme. The self-magnetization ensures gauge invariance and should be expected to play an important role for more sophisticated \textit{ab initio} approaches. We define the dressed photonic frequency $\bar{\omega}_k^2 = \omega_k^2 \big[ 1 + 2\sum_n \sum_{i,j=1}^{3} \chi^m_{n,ij} \tilde{\eps}^{\lambda}_{k,i}(z) \tilde{\eps}^{\lambda}_{k,j}(z) / (c^2 \varepsilon_0 V) \big]$ 
which is related via the sum-rule $\bar{\omega}_k^2 = 2\sum_{n} (E_n-E_m)/\hbar^2 \vert \langle m\vert \hat{p}_{k} \vert n \rangle \vert^2 \quad \forall~m$ to the eigenvalues $E_m$ and eigenstates $\vert m \rangle$. 

In order to provide analytical solutions, we will limit our selves in the following to \textit{either} two-level systems $ \hat{\bmu}_n \rightarrow (\bmu^n_{10} \hat{\sigma}_+ +  \bmu^{n}_{01} \hat{\sigma}_-) $ \textit{or} harmonic oscillators $ \hat{\bmu}_n \rightarrow (\bmu^{n,*} \hat{b}^\dagger + \bmu^{n} \hat{b}) $. Both approaches are widely used but we will focus ultimately on the harmonic representation as it allows to access non-perturbative correlation in the analytic solution. Our molecules are neutral such that we disregard permanent dipole moments for brevity.

The relationship between the transition dipole moments has the generic form $ \textbf{m}^n_{01} = -ic  \dyad{\xi} \bmu^{n}_{01}$, where $\dyad{\xi}$ is the product of a 3D rotation and scaling (see SI). For brevity, we are going to limit our analysis to molecules with collinear transition dipole moments and refer the reader to the SI for a generalization. In this scalar case, $\xi = +1$ and $\xi = -1$ describe an ideal LH and RH emitter, respectively \cite{Condon1937}.
This allows us to combine electric dipole, electric quadrupole and magnetic dipole into a single compact expression when $\boldsymbol\mu^n_{10}=\boldsymbol\mu^n_{01}$ and $Q_{ab,n}^{10}=Q_{ab,n}^{01},~ \textbf{Q}_n = Q_{ab,n}^{10} \nabla_a \textbf{e}_b$. 
%
%
From here on, we are left to follow two different but equally popular directions that we detail in the following. 


A convenient and widely used approximation in quantum optics is to assume that the molecular basis consists of merely two states, motivated by the anharmonicity of excitonic transitions. For very large coupling strength, there is little reason to believe that the complicated multi-level structure of chiral molecules is well captured by only a single excitation. 
Let us assume for a moment we would limit ourselves to this parameter regime. The excitation spectrum is then approximated by a single excitation of energy $\hbar \omega_m$ which is commonly transferred into a Pauli spin-basis $\vert 1 \rangle \langle 0 \vert \rightarrow \sigma^+$. If we further discard the self-polarization term and counter-rotating terms ($\hat{a}\hat{\sigma}^-,~\hat{a}^\dagger\hat{\sigma}^+$), we obtain the strongly simplified chiral Tavis-Cummings Hamiltonian
\begin{align*}
    \hat{H}_{CTC} &= \sum_n \hbar\omega_m \hat{\sigma}^+_n \hat{\sigma}^-_n + \hbar \bar{\omega}_k (\hat{a}^\dagger \hat{a} + \frac{1}{2}) \\
&- i\hbar \sum_n \bar{g}_n (1+\bar{\xi}_n \lambda) \big[ \hat{\sigma}^+_n \hat{a} - \hat{\sigma}^-_n \hat{a}^\dagger\big]\notag~.
\end{align*}
The chiral coupling is encoded via the effective interaction strength proportional to $\bar{g}_n (1+\bar{\xi}_n \lambda)$. A chiral emitter that features the same (opposite) handedness as the cavity will couple stronger (weaker) to the mode. In the extreme case that $\bar{\xi}_n = \pm 1$, the mismatched enantiomer will entirely decouple from the mode. The above chiral Tavis-Cummings model can be solved analogously to the standard Tavis-Cummings model, i.e, by limiting ourselves to the single-excitation subspace and introducing collective spin-operators. 




Once we approach the ultra-strong coupling domain, many excitations of the chiral molecule will contribute to the renormalization of transitions. A possible alternative is the harmonic approximation, in analogy to Hopfield \cite{hopfield1958}, in which we identify the excitation structure with that of an harmonic oscillator. The resulting Hamiltonian takes the form of $N+1$ coupled harmonic oscillators and can be solved via Hopfield diagonalization (detailed in the SI). The Hopfield solution is known to provide accurate predictions for effectively bosonic systems, such as vibrations \cite{george2016} and intersubband transitions \cite{todorov2012}. We find that the qualitative predictions of our Hopfield model are consistent with available \textit{ab initio} calculations (see SI and following). It should be noted that both models provide consistent predictions for the first polariton-manifold under strong coupling as $\hat{\sigma}_{+} \leftrightarrow \hat{b}^{\dagger}$ in the single-excitation space.

Assuming identical (but distinguishable) molecules and homogeneous in-plane distribution, it is convenient to introduce collective molecular operators $\hat{B}_{\bk_{\parallel}}^\dagger = \frac{1}{\sqrt{N}} \sum_{n} e^{i \bk_{\parallel} \cdot \br_n} \hat{b}_n^\dagger,~ \hat{b}_n^\dagger = \frac{1}{\sqrt{N}} \sum_{\bk_{\parallel}} e^{-i \bk_{\parallel} \cdot \br_n} \hat{B}_{\bk_{\parallel}}^\dagger$, where $\bk_{\parallel}$ is the in-plane momentum of the matter excitation. 
The general feature of chiral selectivity is unchanged (see SI).
Under those approximations
\begin{align}
    \hat{H} &\approx  \hbar\widetilde\omega_m \big(\hat{B}_{k=0}^\dagger\hat{B}_{k=0} + \tfrac{1}{2}\big) + \hbar\bar{\omega}_k \big(\ha ^\dagger\ha +\tfrac{1}{2}\big)\notag\\
    &- i \hbar \sqrt{N}\tilde g \big[ (\hat{B}_{k=0}^\dagger + \hat{B}_{k=0}) (\ha  - \ha ^\dagger)     \label{eq:hopfield}\\ 
    &\qquad\qquad\quad+ \tilde \xi \lambda  (\hat{B}_{k=0}^\dagger - \hat{B}_{k=0})(\ha  + \ha ^\dagger) \big]~,\notag
\end{align}
where $ \tilde{\omega}_m^2 = \omega_m^2 + N (2\omega_m/\hbar \varepsilon_0 V) (\tilde{\eps}^{\lambda}_{k}(z) \cdot \bmu)^2 $, 
$\tilde{g} = \sqrt{\hbar \bar{\omega}_k \omega_m/(2\varepsilon_0 V \tilde{\omega}_m)} (\boldsymbol\mu + \textbf{Q}) \cdot \tilde{\eps}^{\lambda}_{k}(z)$ and $\tilde{\xi} = \xi \tilde{\omega}_m \omega_k \boldsymbol\mu\cdot \tilde{\eps}^{\lambda}_{k}(z) /(\omega_m \bar{\omega}_k (\boldsymbol\mu + \textbf{Q}) \cdot \tilde{\eps}^{\lambda}_{k}(z))$ 
are the renormalized effective excitation energy, coupling strength, and chirality factor.

Eq.~\eqref{eq:hopfield} is diagonalized by following the standard Hopfield \cite{hopfield1958,todorov2012} procedure, i.e., defining the polaritonic operator $\hat{\Pi} = x \ha  + y \ha ^\dagger + z \hat{B}_{k=0}  + u \hat{B}_{k=0}^\dagger$  that fulfills the eigenvalue equation $[\hat{H},\hat{\Pi}] = \hbar \Omega \hat{\Pi}$ with the normalization condition $\vert x \vert^2 - \vert y \vert^2 + \vert z \vert^2 - \vert u \vert^2 = 1$.  We obtain the polaritonic frequencies 
as real and positive solutions
\begin{equation}
\begin{split}
    \Omega_{\pm} & = \frac{1}{\sqrt{2}}
    \biggl\{ \bar{\omega}_k^2 + \tilde{\omega}_m^2 + 8 \tilde{\xi}\lambda N \tilde{g}^2 
    \pm \Bigl[ (\bar{\omega}_k^2 - \tilde{\omega}_m^2)^2  \\
    & + 16 N \tilde{g}^2 (\bar{\omega}_k + \tilde\omega_m \tilde\xi \lambda)(\bar{\omega}_k \tilde\xi \lambda + \tilde\omega_m) 
    \Bigr] ^{\frac{1}{2}}
    \biggr\}^{\frac{1}{2}},
\end{split}
\label{Eq_eigenvalues}
\end{equation}
with eigenvalues $E_{\pm}  = \hbar \Omega_{\pm} + E_{vac}, ~ E_{vac} = \hbar (\Omega_{+} + \Omega_{-})/2 $ up to an arbitrary constant that is independent of the handedness of the emitter.

\begin{figure}[t!]
\includegraphics[width=1\columnwidth]{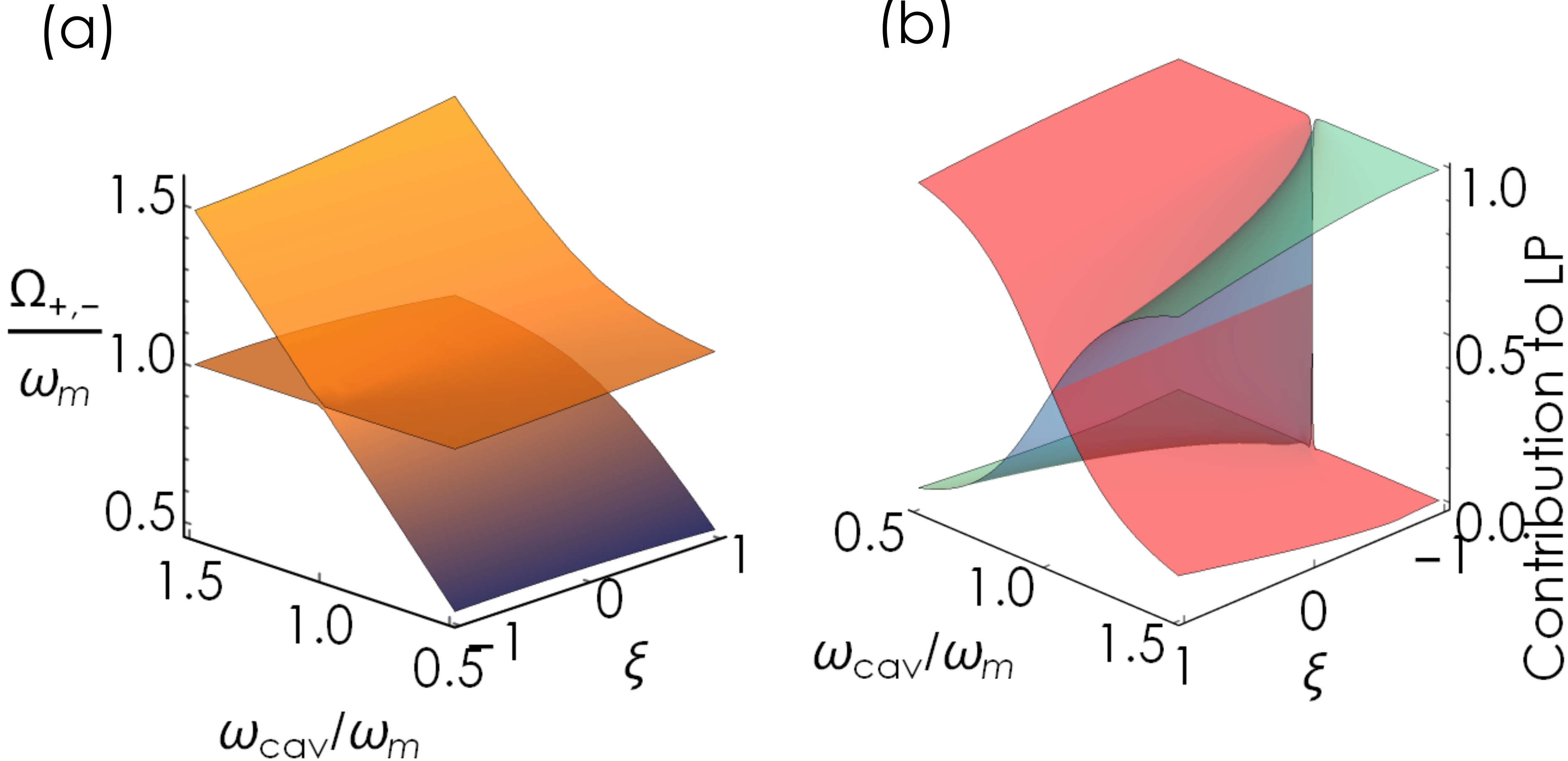}
\caption{(a) Polaritonic eigenvalues $ \Omega_{\pm}$ of the chiral Hopfield model with a LH cavity mode ($\lambda=1$) for $N=100$ as a function of the cavity frequency $\omega_k$ and chiral factor $\xi$. 
The eigenvalues are calculated with typical values for optical transitions in dye molecules \cite{Govorov2010}, $\textbf{Q}=\chi^m_{i,j}=z=0$, and the fundamental coupling strength of $\sqrt{1/\varepsilon_0 V}=0.001$ (a.u.).
(b) Hopfield coefficients (light blue -- photonic, red -- matter) for the lower polariton for the same system as in panel (a).
}
\label{fig:hopfieldscan}
\end{figure}

As illustrated in Fig.~\ref{fig:hopfieldscan}(a), an ensemble of ideal chiral emitters featuring the opposite handedness ($\tilde{\xi} \lambda = -1$) compared to the LH cavity mode effectively decouples from the cavity. Switching the cavity handedness would therefore allow to open and close the avoided crossing and control associated conical intersections.
This is an intuitively expected result: an ensemble of LH molecules couples to the LH photonic mode, whereas an ensemble of equivalent RH molecules becomes transparent for the same optical mode. The cross-section of the full plot at $\xi=0$ yields the familiar picture of "traditional" polaritons with the electric-dipole-mediated coupling \cite{todorov2012}.
Furthermore, the photonic and matter polariton fractions exhibit a gradual transition from the regime of hybridized eigenstates at $\tilde{\xi} \lambda =1$ to the uncoupled regime at $\tilde{\xi} \lambda = -1$, when the two eigenstates represent bare optical and matter excitations.
Our simple Hopfield solution recovers therefore the promising feature that the vacuum coupling in chiral cavities can be used to discriminate between two enantiomers. 
Importantly, the cavity distinguishes only the chiral component, i.e., the parallel projection of $\boldsymbol\mu$. This can be easily seen by extending our discussion to consider general bi-isotropic media and performing an angular average (see SI).

\begin{figure}[t!]
    \centering
    \includegraphics[width=.8\columnwidth]{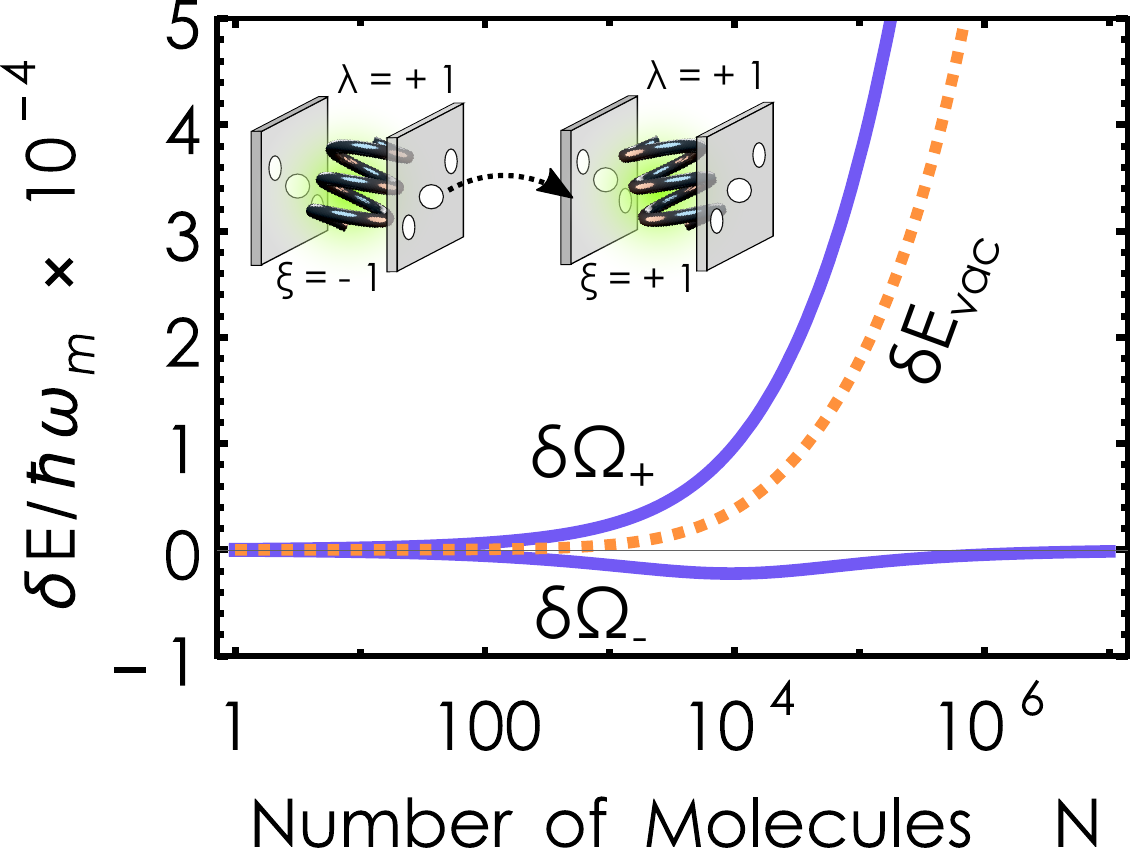}
    \caption{Normalized difference between the polaritonic excitation energies $\delta\Omega_{\pm}=\Omega_{\pm}^{\xi=+1}-\Omega_{\pm}^{\xi=-1}$ (blue-solid) and the correlated ground state $\delta E_{vac} = E_{vac}^{{\xi=+1}}- E_{vac}^{{\xi=-1}}$ (orange-dotted) of left and right-handed chiral dye molecules inside a LH chiral cavity.
    We have chosen common values for dye molecules \cite{Govorov2010}, the resonant condition $\omega_k=\omega_m$ and $\sqrt{1/\varepsilon_0 V}=0.001$ in atomic units for the chiral Hopfield model. For comparison, a molecular concentration of 1 mol/L is on the order of magnitude of $10^3$ molecules in the chosen volume. The actual number of collectively coupled emitters under experimental conditions is usually unknown and is estimated based on simplified models, such as the one presented here.}
    \label{fig:diffeigs}
\end{figure}

The vast majority of widely used molecular systems exhibits extremely weak chirality factors $\xi \ll 1$, rendering it challenging to separate left- and right-handed enantiomers to high fidelity.
Chiral polaritonics can serve this purpose as the collective interaction results in a $\sqrt{N}$ scaling of the coupling strength, thus increasing the selectivity. 
Fig.~\ref{fig:diffeigs} illustrates the difference of upper and lower polaritonic eigenfrequencies (blue) between left- and right-handed enantiomer for typical dye molecules \cite{Govorov2010} with a small and conservative estimate of $\xi \approx 3.712 \cdot 10^{-5} $ for the chirality factor. 

For large $N$, the interacting system enters into the ultra-strong coupling domain in which the combined light-matter ground state is no longer separable.
The relative energy difference between LH/LH and RH/LH ground states is shown in Fig. 3 (orange-dotted) as a function of the number of molecules $N$. Clearly, the correlated ground and excited states illustrate a quick increase of the discriminating effect -- the center-point of chiral polaritonics. The ground-state discrimination scales linear in $N$ for moderate coupling and continues to scale as $\sqrt{N}$ in the deep ultra-strong coupling domain (see SI). Both limits are consistent with the observation by Riso et al.~\cite{riso2022strong}.

The small value of $\xi$ for typical molecules translates into an overall small eigenvalue difference that scales on resonance approximately with $\sqrt{N}\tilde{g}\tilde{\xi}$. 
While $\sqrt{N}\tilde{g}$ can reach a sizeable fraction of the excitation energy, a major limitation to be fought here is the commonly weak chirality. In order to leverage chiral polaritonics for enantiomer selectivity, it seems thus essential to either magnify the magnetic components or establish a protocol that can exploit the small energetic differences. 

The latter follows closely the still open question for the origin of homochirality, i.e., how could a minute energetic imbalance between the enantiomers result in the real-world dominance of a given handedness \cite{doi:10.1126/science.309.5731.78b}. Among the frequently discussed options are autocatalytic processes that turn a small imbalance into a substantial excess \cite{blackmond2004}. Chiral polaritonics could not just explore a similar path but serve as a sensitive framework to further elucidate the origin of homochirality.

The former approach on the other hand would propose the design of a cavity that would compensate for the small $\tilde \xi$. The discriminating factor $1+\tilde\xi \lambda$ is in our purely transversal cavity bound by the small size of $\tilde\xi$ as $\vert \lambda \vert = 1$ is fixed by $Z \bH = - i \lambda \bE$.
The latter, however, no longer holds in subwavelength cavities, where the field has a significant longitudinal component. 
In (non-chiral) plasmonic nanocavities, for example, $|Z\bH| \ll |\bE|$.
Thus, the challenge could be addressed by designing a compact chiral nanocavity, whose quasi-normal mode is dominated by the longitudinal magnetic field, $|Z\bH| \gg |\bE|$, and maintains a chiral character expressed by a non-zero local chirality density $C(\br)$.

We developed a new analytical model describing non-perturbative interaction of an ensemble of chiral molecules with a common chiral optical mode, i.e., a resonator that supports only optical modes with a given handedness. The model illustrates that a chiral cavity can be used to selectively couple to molecules of a specific handedness and thus provides means to discriminate enantiomers from a racemic.
Such a chiral discriminating effect can be observed in all eigenstates of the strongly hybridized light-matter system and exists in absence of any pumping, i.e., in the dark cavity. How strong left and right-handed enantiomer can be distinguished is proportional to $\sqrt{N} g \xi $. While $\sqrt{N} g$ can become sizeable, the degree of chirality satisfies typically $\xi \ll 1$ which currently limits the capability for chiral recognition. Possible strategies to exploit field enhancement techniques \cite{Govorov2010} in combined optical and plasmonic systems \cite{hertzog2021enhancing} might pave a way to enhance the recognition capabilities. It should be noted that this simple, accurately controllable and easily realizable system breaks a discrete symmetry with possibly wide and yet unforeseen ramifications. Chiral polaritonics contributes, even at this early stage, an exciting perspective to further elucidate homochirality.
The intuitive analytical model and perspective put forward in this letter will foster this new domain on the intersection of cavity QED, chiral chemistry, biology and nanophotonics.

\begin{acknowledgement}
We thank G\"oran Johansson, Maxim Gorkunov, and Timur Shegai for stimulating discussions. C.S. acknowledges support from the Swedish Research Council (VR) through Grant No. 2016-06059. D.G.B. acknowledges support from the  Russian Science Foundation
(21-72-00051) and BASIS foundation (Grant No. 22-1-3-2-1).
\end{acknowledgement}

\bibliography{Chirality}

\providecommand{\latin}[1]{#1}
\makeatletter
\providecommand{\doi}
  {\begingroup\let\do\@makeother\dospecials
  \catcode`\{=1 \catcode`\}=2 \doi@aux}
\providecommand{\doi@aux}[1]{\endgroup\texttt{#1}}
\makeatother
\providecommand*\mcitethebibliography{\thebibliography}
\csname @ifundefined\endcsname{endmcitethebibliography}
  {\let\endmcitethebibliography\endthebibliography}{}
\begin{mcitethebibliography}{91}
\providecommand*\natexlab[1]{#1}
\providecommand*\mciteSetBstSublistMode[1]{}
\providecommand*\mciteSetBstMaxWidthForm[2]{}
\providecommand*\mciteBstWouldAddEndPuncttrue
  {\def\EndOfBibitem{\unskip.}}
\providecommand*\mciteBstWouldAddEndPunctfalse
  {\let\EndOfBibitem\relax}
\providecommand*\mciteSetBstMidEndSepPunct[3]{}
\providecommand*\mciteSetBstSublistLabelBeginEnd[3]{}
\providecommand*\EndOfBibitem{}
\mciteSetBstSublistMode{f}
\mciteSetBstMaxWidthForm{subitem}{(\alph{mcitesubitemcount})}
\mciteSetBstSublistLabelBeginEnd
  {\mcitemaxwidthsubitemform\space}
  {\relax}
  {\relax}

\bibitem[Garcia-Vidal \latin{et~al.}(2021)Garcia-Vidal, Ciuti, and
  Ebbesen]{garcia2021manipulating}
Garcia-Vidal,~F.~J.; Ciuti,~C.; Ebbesen,~T.~W. Manipulating matter by strong
  coupling to vacuum fields. \emph{Science} \textbf{2021}, \emph{373},
  eabd0336\relax
\mciteBstWouldAddEndPuncttrue
\mciteSetBstMidEndSepPunct{\mcitedefaultmidpunct}
{\mcitedefaultendpunct}{\mcitedefaultseppunct}\relax
\EndOfBibitem
\bibitem[Simpkins \latin{et~al.}(2021)Simpkins, Dunkelberger, and
  Owrutsky]{simpkins2021mode}
Simpkins,~B.~S.; Dunkelberger,~A.~D.; Owrutsky,~J.~C. Mode-specific chemistry
  through vibrational strong coupling (or A wish come true). \emph{The Journal
  of Physical Chemistry C} \textbf{2021}, \emph{125}, 19081--19087\relax
\mciteBstWouldAddEndPuncttrue
\mciteSetBstMidEndSepPunct{\mcitedefaultmidpunct}
{\mcitedefaultendpunct}{\mcitedefaultseppunct}\relax
\EndOfBibitem
\bibitem[Sidler \latin{et~al.}(2022)Sidler, Ruggenthaler, Sch{\"a}fer, Ronca,
  and Rubio]{sidler2021perspective}
Sidler,~D.; Ruggenthaler,~M.; Sch{\"a}fer,~C.; Ronca,~E.; Rubio,~A. A
  perspective on ab initio modeling of polaritonic chemistry: The role of
  non-equilibrium effects and quantum collectivity. \emph{The Journal of
  Chemical Physics} \textbf{2022}, \emph{156}, 230901\relax
\mciteBstWouldAddEndPuncttrue
\mciteSetBstMidEndSepPunct{\mcitedefaultmidpunct}
{\mcitedefaultendpunct}{\mcitedefaultseppunct}\relax
\EndOfBibitem
\bibitem[Forn-D{\'\i}az \latin{et~al.}(2019)Forn-D{\'\i}az, Lamata, Rico, Kono,
  and Solano]{forn2019ultrastrong}
Forn-D{\'\i}az,~P.; Lamata,~L.; Rico,~E.; Kono,~J.; Solano,~E. Ultrastrong
  coupling regimes of light-matter interaction. \emph{Rev. Mod. Phys.}
  \textbf{2019}, \emph{91}, 025005\relax
\mciteBstWouldAddEndPuncttrue
\mciteSetBstMidEndSepPunct{\mcitedefaultmidpunct}
{\mcitedefaultendpunct}{\mcitedefaultseppunct}\relax
\EndOfBibitem
\bibitem[Kockum \latin{et~al.}(2019)Kockum, Miranowicz, De~Liberato, Savasta,
  and Nori]{kockum2019ultrastrong}
Kockum,~A.~F.; Miranowicz,~A.; De~Liberato,~S.; Savasta,~S.; Nori,~F.
  Ultrastrong coupling between light and matter. \emph{Nat. Rev. Phys.}
  \textbf{2019}, \emph{1}, 19\relax
\mciteBstWouldAddEndPuncttrue
\mciteSetBstMidEndSepPunct{\mcitedefaultmidpunct}
{\mcitedefaultendpunct}{\mcitedefaultseppunct}\relax
\EndOfBibitem
\bibitem[Mennucci and Corni(2019)Mennucci, and Corni]{mennucci2019multiscale}
Mennucci,~B.; Corni,~S. Multiscale modelling of photoinduced processes in
  composite systems. \emph{Nature Reviews Chemistry} \textbf{2019}, \emph{3},
  315--330\relax
\mciteBstWouldAddEndPuncttrue
\mciteSetBstMidEndSepPunct{\mcitedefaultmidpunct}
{\mcitedefaultendpunct}{\mcitedefaultseppunct}\relax
\EndOfBibitem
\bibitem[Luk \latin{et~al.}(2017)Luk, Feist, Toppari, and
  Groenhof]{luk2017multiscale}
Luk,~H.~L.; Feist,~J.; Toppari,~J.~J.; Groenhof,~G. Multiscale Molecular
  Dynamics Simulations of Polaritonic Chemistry. \emph{Journal of chemical
  theory and computation} \textbf{2017}, \emph{13}, 4324--4335\relax
\mciteBstWouldAddEndPuncttrue
\mciteSetBstMidEndSepPunct{\mcitedefaultmidpunct}
{\mcitedefaultendpunct}{\mcitedefaultseppunct}\relax
\EndOfBibitem
\bibitem[Fregoni \latin{et~al.}(2021)Fregoni, Haugland, Pipolo, Giovannini,
  Koch, and Corni]{fregoni2021strong}
Fregoni,~J.; Haugland,~T.~S.; Pipolo,~S.; Giovannini,~T.; Koch,~H.; Corni,~S.
  Strong coupling between localized surface plasmons and molecules by coupled
  cluster theory. \emph{Nano Letters} \textbf{2021}, \emph{21},
  6664--6670\relax
\mciteBstWouldAddEndPuncttrue
\mciteSetBstMidEndSepPunct{\mcitedefaultmidpunct}
{\mcitedefaultendpunct}{\mcitedefaultseppunct}\relax
\EndOfBibitem
\bibitem[Bajoni \latin{et~al.}(2008)Bajoni, Senellart, Wertz, Sagnes, Miard,
  Lema{\^\i}tre, and Bloch]{bajoni2008polariton}
Bajoni,~D.; Senellart,~P.; Wertz,~E.; Sagnes,~I.; Miard,~A.; Lema{\^\i}tre,~A.;
  Bloch,~J. Polariton laser using single micropillar GaAs- GaAlAs semiconductor
  cavities. \emph{Physical review letters} \textbf{2008}, \emph{100},
  047401\relax
\mciteBstWouldAddEndPuncttrue
\mciteSetBstMidEndSepPunct{\mcitedefaultmidpunct}
{\mcitedefaultendpunct}{\mcitedefaultseppunct}\relax
\EndOfBibitem
\bibitem[Chikkaraddy \latin{et~al.}(2016)Chikkaraddy, de~Nijs, Benz, Barrow,
  Scherman, Rosta, Demetriadou, Fox, Hess, and Baumberg]{chikkaraddy2016}
Chikkaraddy,~R.; de~Nijs,~B.; Benz,~F.; Barrow,~S.~J.; Scherman,~O.~A.;
  Rosta,~E.; Demetriadou,~A.; Fox,~P.; Hess,~O.; Baumberg,~J.~J.
  Single-molecule strong coupling at room temperature in plasmonic
  nanocavities. \emph{Nature} \textbf{2016}, \emph{535}, 127--130\relax
\mciteBstWouldAddEndPuncttrue
\mciteSetBstMidEndSepPunct{\mcitedefaultmidpunct}
{\mcitedefaultendpunct}{\mcitedefaultseppunct}\relax
\EndOfBibitem
\bibitem[Wang \latin{et~al.}(2017)Wang, Kelkar, Martin-Cano, Utikal,
  G\"otzinger, and Sandoghdar]{wang2017coherent}
Wang,~D.; Kelkar,~H.; Martin-Cano,~D.; Utikal,~T.; G\"otzinger,~S.;
  Sandoghdar,~V. Coherent Coupling of a Single Molecule to a Scanning
  Fabry-Perot Microcavity. \emph{Phys. Rev. X} \textbf{2017}, \emph{7},
  021014\relax
\mciteBstWouldAddEndPuncttrue
\mciteSetBstMidEndSepPunct{\mcitedefaultmidpunct}
{\mcitedefaultendpunct}{\mcitedefaultseppunct}\relax
\EndOfBibitem
\bibitem[Baranov \latin{et~al.}(2020)Baranov, Munkhbat, Zhukova, Bisht,
  Canales, Rousseaux, Johansson, Antosiewicz, and
  Shegai]{baranov2019ultrastrong}
Baranov,~D.~G.; Munkhbat,~B.; Zhukova,~E.; Bisht,~A.; Canales,~A.;
  Rousseaux,~B.; Johansson,~G.; Antosiewicz,~T.~J.; Shegai,~T. Ultrastrong
  coupling between nanoparticle plasmons and cavity photons at ambient
  conditions. \emph{Nat. Commun.} \textbf{2020}, \emph{11}, 2715\relax
\mciteBstWouldAddEndPuncttrue
\mciteSetBstMidEndSepPunct{\mcitedefaultmidpunct}
{\mcitedefaultendpunct}{\mcitedefaultseppunct}\relax
\EndOfBibitem
\bibitem[H{\"u}bener \latin{et~al.}(2021)H{\"u}bener, De~Giovannini,
  Sch{\"a}fer, Andberger, Ruggenthaler, Faist, and
  Rubio]{hubener2021engineering}
H{\"u}bener,~H.; De~Giovannini,~U.; Sch{\"a}fer,~C.; Andberger,~J.;
  Ruggenthaler,~M.; Faist,~J.; Rubio,~A. Engineering quantum materials with
  chiral optical cavities. \emph{Nature materials} \textbf{2021}, \emph{20},
  438--442\relax
\mciteBstWouldAddEndPuncttrue
\mciteSetBstMidEndSepPunct{\mcitedefaultmidpunct}
{\mcitedefaultendpunct}{\mcitedefaultseppunct}\relax
\EndOfBibitem
\bibitem[Gubbin and De~Liberato(2020)Gubbin, and De~Liberato]{gubbin2020}
Gubbin,~C.~R.; De~Liberato,~S. Optical Nonlocality in Polar Dielectrics.
  \emph{Phys. Rev. X} \textbf{2020}, \emph{10}, 021027\relax
\mciteBstWouldAddEndPuncttrue
\mciteSetBstMidEndSepPunct{\mcitedefaultmidpunct}
{\mcitedefaultendpunct}{\mcitedefaultseppunct}\relax
\EndOfBibitem
\bibitem[Thomas \latin{et~al.}(2021)Thomas, Menghrajani, and
  Barnes]{doi:10.1021/acs.jpclett.1c01695}
Thomas,~P.~A.; Menghrajani,~K.~S.; Barnes,~W.~L. Cavity-Free Ultrastrong
  Light-Matter Coupling. \emph{The Journal of Physical Chemistry Letters}
  \textbf{2021}, \emph{12}, 6914--6918\relax
\mciteBstWouldAddEndPuncttrue
\mciteSetBstMidEndSepPunct{\mcitedefaultmidpunct}
{\mcitedefaultendpunct}{\mcitedefaultseppunct}\relax
\EndOfBibitem
\bibitem[Coles \latin{et~al.}(2014)Coles, Somaschi, Michetti, Clark,
  Lagoudakis, Savvidis, and Lidzey]{coles2014b}
Coles,~D.~M.; Somaschi,~N.; Michetti,~P.; Clark,~C.; Lagoudakis,~P.~G.;
  Savvidis,~P.~G.; Lidzey,~D.~G. Polariton-mediated energy transfer between
  organic dyes in a strongly coupled optical microcavity. \emph{Nat. Mater.}
  \textbf{2014}, \emph{13}, 712--719\relax
\mciteBstWouldAddEndPuncttrue
\mciteSetBstMidEndSepPunct{\mcitedefaultmidpunct}
{\mcitedefaultendpunct}{\mcitedefaultseppunct}\relax
\EndOfBibitem
\bibitem[Orgiu \latin{et~al.}(2015)Orgiu, George, Hutchison, Devaux, Dayen,
  Doudin, Stellacci, Genet, Schachenmayer, Genes, Pupillo, Samor{\`{\i}}, and
  Ebbesen]{orgiu2015}
Orgiu,~E.; George,~J.; Hutchison,~J.~A.; Devaux,~E.; Dayen,~J.~F.; Doudin,~B.;
  Stellacci,~F.; Genet,~C.; Schachenmayer,~J.; Genes,~C. \latin{et~al.}
  Conductivity in organic semiconductors hybridized with the vacuum field.
  \emph{Nat. Mater.} \textbf{2015}, \emph{14}, 1123--1129\relax
\mciteBstWouldAddEndPuncttrue
\mciteSetBstMidEndSepPunct{\mcitedefaultmidpunct}
{\mcitedefaultendpunct}{\mcitedefaultseppunct}\relax
\EndOfBibitem
\bibitem[Zhong \latin{et~al.}(2017)Zhong, Chervy, Zhang, Thomas, George, Genet,
  Hutchison, and Ebbesen]{zhong2017energy}
Zhong,~X.; Chervy,~T.; Zhang,~L.; Thomas,~A.; George,~J.; Genet,~C.;
  Hutchison,~J.~A.; Ebbesen,~T.~W. Energy Transfer between Spatially Separated
  Entangled Molecules. \emph{Angew. Chem. Int. Ed.} \textbf{2017}, \emph{56},
  9034--9038\relax
\mciteBstWouldAddEndPuncttrue
\mciteSetBstMidEndSepPunct{\mcitedefaultmidpunct}
{\mcitedefaultendpunct}{\mcitedefaultseppunct}\relax
\EndOfBibitem
\bibitem[Sch{\"a}fer \latin{et~al.}(2019)Sch{\"a}fer, Ruggenthaler, Appel, and
  Rubio]{schafer2019modification}
Sch{\"a}fer,~C.; Ruggenthaler,~M.; Appel,~H.; Rubio,~A. Modification of
  excitation and charge transfer in cavity quantum-electrodynamical chemistry.
  \emph{Proceedings of the National Academy of Sciences} \textbf{2019},
  \emph{116}, 4883--4892\relax
\mciteBstWouldAddEndPuncttrue
\mciteSetBstMidEndSepPunct{\mcitedefaultmidpunct}
{\mcitedefaultendpunct}{\mcitedefaultseppunct}\relax
\EndOfBibitem
\bibitem[Du \latin{et~al.}(2018)Du, Mart{\'\i}nez-Mart{\'\i}nez, Ribeiro, Hu,
  Menon, and Yuen-Zhou]{du2018theory}
Du,~M.; Mart{\'\i}nez-Mart{\'\i}nez,~L.~A.; Ribeiro,~R.~F.; Hu,~Z.;
  Menon,~V.~M.; Yuen-Zhou,~J. Theory for polariton-assisted remote energy
  transfer. \emph{Chem. Sci.} \textbf{2018}, \emph{9}, 6659--6669\relax
\mciteBstWouldAddEndPuncttrue
\mciteSetBstMidEndSepPunct{\mcitedefaultmidpunct}
{\mcitedefaultendpunct}{\mcitedefaultseppunct}\relax
\EndOfBibitem
\bibitem[Hagenm\"uller \latin{et~al.}(2018)Hagenm\"uller, Sch\"utz,
  Schachenmayer, Genes, and Pupillo]{hagenmuller2018}
Hagenm\"uller,~D.; Sch\"utz,~S.; Schachenmayer,~J.; Genes,~C.; Pupillo,~G.
  Cavity-assisted mesoscopic transport of fermions: Coherent and dissipative
  dynamics. \emph{Phys. Rev. B} \textbf{2018}, \emph{97}, 205303\relax
\mciteBstWouldAddEndPuncttrue
\mciteSetBstMidEndSepPunct{\mcitedefaultmidpunct}
{\mcitedefaultendpunct}{\mcitedefaultseppunct}\relax
\EndOfBibitem
\bibitem[Cohn \latin{et~al.}(2022)Cohn, Sufrin, Basu, and Chuntonov]{cohn2022}
Cohn,~B.; Sufrin,~S.; Basu,~A.; Chuntonov,~L. Vibrational Polaritons in
  Disordered Molecular Ensembles. \emph{The Journal of Physical Chemistry
  Letters} \textbf{2022}, \emph{13}, 8369--8375\relax
\mciteBstWouldAddEndPuncttrue
\mciteSetBstMidEndSepPunct{\mcitedefaultmidpunct}
{\mcitedefaultendpunct}{\mcitedefaultseppunct}\relax
\EndOfBibitem
\bibitem[Hutchison \latin{et~al.}(2012)Hutchison, Schwartz, Genet, Devaux, and
  Ebbesen]{hutchison2012}
Hutchison,~J.~A.; Schwartz,~T.; Genet,~C.; Devaux,~E.; Ebbesen,~T.~W. Modifying
  Chemical Landscapes by Coupling to Vacuum Fields. \emph{Angew. Chem. Int.
  Ed.} \textbf{2012}, \emph{51}, 1592--1596\relax
\mciteBstWouldAddEndPuncttrue
\mciteSetBstMidEndSepPunct{\mcitedefaultmidpunct}
{\mcitedefaultendpunct}{\mcitedefaultseppunct}\relax
\EndOfBibitem
\bibitem[Thomas \latin{et~al.}(2019)Thomas, Lethuillier-Karl, Nagarajan,
  Vergauwe, George, Chervy, Shalabney, Devaux, Genet, Moran, \latin{et~al.}
  others]{thomas2019tilting}
Thomas,~A.; Lethuillier-Karl,~L.; Nagarajan,~K.; Vergauwe,~R.~M.; George,~J.;
  Chervy,~T.; Shalabney,~A.; Devaux,~E.; Genet,~C.; Moran,~J. \latin{et~al.}
  Tilting a ground-state reactivity landscape by vibrational strong coupling.
  \emph{Science} \textbf{2019}, \emph{363}, 615--619\relax
\mciteBstWouldAddEndPuncttrue
\mciteSetBstMidEndSepPunct{\mcitedefaultmidpunct}
{\mcitedefaultendpunct}{\mcitedefaultseppunct}\relax
\EndOfBibitem
\bibitem[Singh \latin{et~al.}(2022)Singh, Lather, and George]{singh2022solvent}
Singh,~J.; Lather,~J.; George,~J. Solvent Dependence on Cooperative Vibrational
  Strong Coupling and Cavity Catalysis. \emph{ChemRxiv} \textbf{2022}, \relax
\mciteBstWouldAddEndPunctfalse
\mciteSetBstMidEndSepPunct{\mcitedefaultmidpunct}
{}{\mcitedefaultseppunct}\relax
\EndOfBibitem
\bibitem[Imperatore \latin{et~al.}(2021)Imperatore, Asbury, and
  Giebink]{imperatore2021reproducibility}
Imperatore,~M.~V.; Asbury,~J.~B.; Giebink,~N.~C. Reproducibility of
  cavity-enhanced chemical reaction rates in the vibrational strong coupling
  regime. \emph{The Journal of Chemical Physics} \textbf{2021}, \emph{154},
  191103\relax
\mciteBstWouldAddEndPuncttrue
\mciteSetBstMidEndSepPunct{\mcitedefaultmidpunct}
{\mcitedefaultendpunct}{\mcitedefaultseppunct}\relax
\EndOfBibitem
\bibitem[Sch{\"a}fer \latin{et~al.}(2022)Sch{\"a}fer, Flick, Ronca, Narang, and
  Rubio]{schafer2021shining}
Sch{\"a}fer,~C.; Flick,~J.; Ronca,~E.; Narang,~P.; Rubio,~A. Shining light on
  the microscopic resonant mechanism responsible for cavity-mediated chemical
  reactivity. \emph{Nature Communications} \textbf{2022}, \emph{13}, 1--9\relax
\mciteBstWouldAddEndPuncttrue
\mciteSetBstMidEndSepPunct{\mcitedefaultmidpunct}
{\mcitedefaultendpunct}{\mcitedefaultseppunct}\relax
\EndOfBibitem
\bibitem[Schäfer(2022)]{schafer2022emb}
Schäfer,~C. Polaritonic Chemistry from First Principles via Embedding
  Radiation Reaction. \emph{The Journal of Physical Chemistry Letters}
  \textbf{2022}, \emph{13}, 6905--6911\relax
\mciteBstWouldAddEndPuncttrue
\mciteSetBstMidEndSepPunct{\mcitedefaultmidpunct}
{\mcitedefaultendpunct}{\mcitedefaultseppunct}\relax
\EndOfBibitem
\bibitem[Li \latin{et~al.}(2021)Li, Nitzan, and Subotnik]{li2021collective}
Li,~T.~E.; Nitzan,~A.; Subotnik,~J.~E. Collective vibrational strong coupling
  effects on molecular vibrational relaxation and energy transfer: Numerical
  insights via cavity molecular dynamics simulations. \emph{Angewandte Chemie}
  \textbf{2021}, \emph{133}, 15661--15668\relax
\mciteBstWouldAddEndPuncttrue
\mciteSetBstMidEndSepPunct{\mcitedefaultmidpunct}
{\mcitedefaultendpunct}{\mcitedefaultseppunct}\relax
\EndOfBibitem
\bibitem[Li \latin{et~al.}(2021)Li, Mandal, and Huo]{li2021cavity}
Li,~X.; Mandal,~A.; Huo,~P. Cavity frequency-dependent theory for vibrational
  polariton chemistry. \emph{Nat. Commun.} \textbf{2021}, \emph{12}, 1315\relax
\mciteBstWouldAddEndPuncttrue
\mciteSetBstMidEndSepPunct{\mcitedefaultmidpunct}
{\mcitedefaultendpunct}{\mcitedefaultseppunct}\relax
\EndOfBibitem
\bibitem[Galego \latin{et~al.}(2016)Galego, Garcia-Vidal, and
  Feist]{galego2016}
Galego,~J.; Garcia-Vidal,~F.~J.; Feist,~J. Suppressing photochemical reactions
  with quantized light fields. \emph{Nat. Commun.} \textbf{2016}, \emph{7},
  13841\relax
\mciteBstWouldAddEndPuncttrue
\mciteSetBstMidEndSepPunct{\mcitedefaultmidpunct}
{\mcitedefaultendpunct}{\mcitedefaultseppunct}\relax
\EndOfBibitem
\bibitem[Munkhbat \latin{et~al.}(2018)Munkhbat, Wers{\"a}ll, Baranov,
  Antosiewicz, and Shegai]{munkhbat2018suppression}
Munkhbat,~B.; Wers{\"a}ll,~M.; Baranov,~D.~G.; Antosiewicz,~T.~J.; Shegai,~T.
  Suppression of photo-oxidation of organic chromophores by strong coupling to
  plasmonic nanoantennas. \emph{Sci. Adv.} \textbf{2018}, \emph{4},
  eaas9552\relax
\mciteBstWouldAddEndPuncttrue
\mciteSetBstMidEndSepPunct{\mcitedefaultmidpunct}
{\mcitedefaultendpunct}{\mcitedefaultseppunct}\relax
\EndOfBibitem
\bibitem[Groenhof \latin{et~al.}(2019)Groenhof, Climent, Feist, Morozov, and
  Toppari]{groenhof2019tracking}
Groenhof,~G.; Climent,~C.; Feist,~J.; Morozov,~D.; Toppari,~J.~J. Tracking
  Polariton Relaxation with Multiscale Molecular Dynamics Simulations. \emph{J.
  Phys. Chem. Lett.} \textbf{2019}, \emph{10}, 5476--5483\relax
\mciteBstWouldAddEndPuncttrue
\mciteSetBstMidEndSepPunct{\mcitedefaultmidpunct}
{\mcitedefaultendpunct}{\mcitedefaultseppunct}\relax
\EndOfBibitem
\bibitem[Kowalewski \latin{et~al.}(2016)Kowalewski, Bennett, and
  Mukamel]{kowalewski2016cavity}
Kowalewski,~M.; Bennett,~K.; Mukamel,~S. Cavity femtochemistry: Manipulating
  nonadiabatic dynamics at avoided crossings. \emph{The journal of physical
  chemistry letters} \textbf{2016}, \emph{7}, 2050--2054\relax
\mciteBstWouldAddEndPuncttrue
\mciteSetBstMidEndSepPunct{\mcitedefaultmidpunct}
{\mcitedefaultendpunct}{\mcitedefaultseppunct}\relax
\EndOfBibitem
\bibitem[Vendrell(2018)]{Vendrell2018}
Vendrell,~O. Collective Jahn-Teller Interactions through Light-Matter Coupling
  in a Cavity. \emph{Phys. Rev. Lett.} \textbf{2018}, \emph{121}, 253001\relax
\mciteBstWouldAddEndPuncttrue
\mciteSetBstMidEndSepPunct{\mcitedefaultmidpunct}
{\mcitedefaultendpunct}{\mcitedefaultseppunct}\relax
\EndOfBibitem
\bibitem[F{\'a}bri \latin{et~al.}(2021)F{\'a}bri, Hal{\'a}sz, and
  Vib{\'o}k]{fabri2021probing}
F{\'a}bri,~C.; Hal{\'a}sz,~G.~J.; Vib{\'o}k,~{\'A}. Probing Light-Induced
  Conical Intersections by Monitoring Multidimensional Polaritonic Surfaces.
  \emph{The Journal of Physical Chemistry Letters} \textbf{2021}, \emph{13},
  1172--1179\relax
\mciteBstWouldAddEndPuncttrue
\mciteSetBstMidEndSepPunct{\mcitedefaultmidpunct}
{\mcitedefaultendpunct}{\mcitedefaultseppunct}\relax
\EndOfBibitem
\bibitem[Li \latin{et~al.}(2022)Li, Tao, and
  Hammes-Schiffer]{li2022semiclassical}
Li,~T.~E.; Tao,~Z.; Hammes-Schiffer,~S. Semiclassical Real-Time
  Nuclear-Electronic Orbital Dynamics for Molecular Polaritons: Unified Theory
  of Electronic and Vibrational Strong Couplings. \emph{Journal of Chemical
  Theory and Computation} \textbf{2022}, \emph{18}, 2774--2784\relax
\mciteBstWouldAddEndPuncttrue
\mciteSetBstMidEndSepPunct{\mcitedefaultmidpunct}
{\mcitedefaultendpunct}{\mcitedefaultseppunct}\relax
\EndOfBibitem
\bibitem[Fischer \latin{et~al.}(2022)Fischer, Anders, and
  Saalfrank]{fischer2022cavity}
Fischer,~E.~W.; Anders,~J.; Saalfrank,~P. Cavity-altered thermal isomerization
  rates and dynamical resonant localization in vibro-polaritonic chemistry.
  \emph{The Journal of Chemical Physics} \textbf{2022}, \emph{156},
  154305\relax
\mciteBstWouldAddEndPuncttrue
\mciteSetBstMidEndSepPunct{\mcitedefaultmidpunct}
{\mcitedefaultendpunct}{\mcitedefaultseppunct}\relax
\EndOfBibitem
\bibitem[Deng \latin{et~al.}(2003)Deng, Weihs, Snoke, Bloch, and
  Yamamoto]{deng2003polariton}
Deng,~H.; Weihs,~G.; Snoke,~D.; Bloch,~J.; Yamamoto,~Y. Polariton lasing vs.
  photon lasing in a semiconductor microcavity. \emph{Proceedings of the
  National Academy of Sciences} \textbf{2003}, \emph{100}, 15318--15323\relax
\mciteBstWouldAddEndPuncttrue
\mciteSetBstMidEndSepPunct{\mcitedefaultmidpunct}
{\mcitedefaultendpunct}{\mcitedefaultseppunct}\relax
\EndOfBibitem
\bibitem[K{\'e}na-Cohen and Forrest(2010)K{\'e}na-Cohen, and
  Forrest]{kena2010room}
K{\'e}na-Cohen,~S.; Forrest,~S. Room-temperature polariton lasing in an organic
  single-crystal microcavity. \emph{Nature Photonics} \textbf{2010}, \emph{4},
  371--375\relax
\mciteBstWouldAddEndPuncttrue
\mciteSetBstMidEndSepPunct{\mcitedefaultmidpunct}
{\mcitedefaultendpunct}{\mcitedefaultseppunct}\relax
\EndOfBibitem
\bibitem[Slootsky \latin{et~al.}(2014)Slootsky, Liu, Menon, and
  Forrest]{slootsky2014room}
Slootsky,~M.; Liu,~X.; Menon,~V.~M.; Forrest,~S.~R. Room temperature
  Frenkel-Wannier-Mott hybridization of degenerate excitons in a strongly
  coupled microcavity. \emph{Phys. Rev. Lett.} \textbf{2014}, \emph{112},
  076401\relax
\mciteBstWouldAddEndPuncttrue
\mciteSetBstMidEndSepPunct{\mcitedefaultmidpunct}
{\mcitedefaultendpunct}{\mcitedefaultseppunct}\relax
\EndOfBibitem
\bibitem[Latini \latin{et~al.}(2021)Latini, Shin, Sato, Sch{\"a}fer,
  De~Giovannini, H{\"u}bener, and Rubio]{latini2021ferroelectric}
Latini,~S.; Shin,~D.; Sato,~S.~A.; Sch{\"a}fer,~C.; De~Giovannini,~U.;
  H{\"u}bener,~H.; Rubio,~A. The ferroelectric photo ground state of
  {SrTiO}$_3$: Cavity materials engineering. \emph{Proceedings of the National
  Academy of Sciences} \textbf{2021}, \emph{118}, e2105618118\relax
\mciteBstWouldAddEndPuncttrue
\mciteSetBstMidEndSepPunct{\mcitedefaultmidpunct}
{\mcitedefaultendpunct}{\mcitedefaultseppunct}\relax
\EndOfBibitem
\bibitem[Flick \latin{et~al.}(2019)Flick, Welakuh, Ruggenthaler, Appel, and
  Rubio]{flick2018light}
Flick,~J.; Welakuh,~D.~M.; Ruggenthaler,~M.; Appel,~H.; Rubio,~A. Light--Matter
  Response in Nonrelativistic Quantum Electrodynamics. \emph{ACS Photonics}
  \textbf{2019}, \emph{6}, 2757--2778\relax
\mciteBstWouldAddEndPuncttrue
\mciteSetBstMidEndSepPunct{\mcitedefaultmidpunct}
{\mcitedefaultendpunct}{\mcitedefaultseppunct}\relax
\EndOfBibitem
\bibitem[Schlawin \latin{et~al.}(2022)Schlawin, Kennes, and
  Sentef]{schlawin2022cavity}
Schlawin,~F.; Kennes,~D.~M.; Sentef,~M.~A. Cavity quantum materials.
  \emph{Applied Physics Reviews} \textbf{2022}, \emph{9}, 011312\relax
\mciteBstWouldAddEndPuncttrue
\mciteSetBstMidEndSepPunct{\mcitedefaultmidpunct}
{\mcitedefaultendpunct}{\mcitedefaultseppunct}\relax
\EndOfBibitem
\bibitem[Sch\"afer and Johansson(2022)Sch\"afer, and
  Johansson]{schafer2021shortcut}
Sch\"afer,~C.; Johansson,~G. Shortcut to Self-Consistent Light-Matter
  Interaction and Realistic Spectra from First Principles. \emph{Phys. Rev.
  Lett.} \textbf{2022}, \emph{128}, 156402\relax
\mciteBstWouldAddEndPuncttrue
\mciteSetBstMidEndSepPunct{\mcitedefaultmidpunct}
{\mcitedefaultendpunct}{\mcitedefaultseppunct}\relax
\EndOfBibitem
\bibitem[Lentrodt \latin{et~al.}(2020)Lentrodt, Heeg, Keitel, and
  Evers]{lentrodt2020}
Lentrodt,~D.; Heeg,~K.~P.; Keitel,~C.~H.; Evers,~J. Ab initio quantum models
  for thin-film x-ray cavity QED. \emph{Phys. Rev. Research} \textbf{2020},
  \emph{2}, 023396\relax
\mciteBstWouldAddEndPuncttrue
\mciteSetBstMidEndSepPunct{\mcitedefaultmidpunct}
{\mcitedefaultendpunct}{\mcitedefaultseppunct}\relax
\EndOfBibitem
\bibitem[Debnath and Rubio(2020)Debnath, and Rubio]{doi:10.1063/5.0012754}
Debnath,~A.; Rubio,~A. Entangled photon assisted multidimensional nonlinear
  optics of exciton–polaritons. \emph{Journal of Applied Physics}
  \textbf{2020}, \emph{128}, 113102\relax
\mciteBstWouldAddEndPuncttrue
\mciteSetBstMidEndSepPunct{\mcitedefaultmidpunct}
{\mcitedefaultendpunct}{\mcitedefaultseppunct}\relax
\EndOfBibitem
\bibitem[Lindell \latin{et~al.}(2018)Lindell, Sihvola, Tretyakov, and
  Vitanen]{Lindell2018}
Lindell,~I.; Sihvola,~A.; Tretyakov,~S.; Vitanen,~A. \emph{{Electromagnetic
  waves in chiral and Bi-isotropic media}}; Artech House, 2018; p 332\relax
\mciteBstWouldAddEndPuncttrue
\mciteSetBstMidEndSepPunct{\mcitedefaultmidpunct}
{\mcitedefaultendpunct}{\mcitedefaultseppunct}\relax
\EndOfBibitem
\bibitem[Barron(2004)]{Barron2004}
Barron,~L.~D. \emph{Molecular Light Scattering and Optical Activity}; Cambridge
  University Press, 2004; p 443\relax
\mciteBstWouldAddEndPuncttrue
\mciteSetBstMidEndSepPunct{\mcitedefaultmidpunct}
{\mcitedefaultendpunct}{\mcitedefaultseppunct}\relax
\EndOfBibitem
\bibitem[Condon(1937)]{Condon1937}
Condon,~E.~U. {Theories of Optical Rotatory Power}. \emph{Reviews of Modern
  Physics} \textbf{1937}, \emph{9}, 432--457\relax
\mciteBstWouldAddEndPuncttrue
\mciteSetBstMidEndSepPunct{\mcitedefaultmidpunct}
{\mcitedefaultendpunct}{\mcitedefaultseppunct}\relax
\EndOfBibitem
\bibitem[Kelvin(1894)]{Kelvin1894}
Kelvin,~W. T.~B. \emph{The Molecular Tactics of a Crystal}; Robert Boyle
  lecture; Clarendon Press, 1894\relax
\mciteBstWouldAddEndPuncttrue
\mciteSetBstMidEndSepPunct{\mcitedefaultmidpunct}
{\mcitedefaultendpunct}{\mcitedefaultseppunct}\relax
\EndOfBibitem
\bibitem[Weiskopf \latin{et~al.}(2002)Weiskopf, Nau, and
  Strichartz]{weiskopf2002}
Weiskopf,~R.~B.; Nau,~C.; Strichartz,~G.~R. {Drug Chirality in Anesthesia}.
  \emph{Anesthesiology} \textbf{2002}, \emph{97}, 497--502\relax
\mciteBstWouldAddEndPuncttrue
\mciteSetBstMidEndSepPunct{\mcitedefaultmidpunct}
{\mcitedefaultendpunct}{\mcitedefaultseppunct}\relax
\EndOfBibitem
\bibitem[Calcaterra and D’Acquarica(2018)Calcaterra, and
  D’Acquarica]{CALCATERRA2018323}
Calcaterra,~A.; D’Acquarica,~I. The market of chiral drugs: Chiral switches
  versus de novo enantiomerically pure compounds. \emph{Journal of
  Pharmaceutical and Biomedical Analysis} \textbf{2018}, \emph{147},
  323--340\relax
\mciteBstWouldAddEndPuncttrue
\mciteSetBstMidEndSepPunct{\mcitedefaultmidpunct}
{\mcitedefaultendpunct}{\mcitedefaultseppunct}\relax
\EndOfBibitem
\bibitem[Scriba(2016)]{SCRIBA201656}
Scriba,~G.~K. Chiral recognition in separation science – an update.
  \emph{Journal of Chromatography A} \textbf{2016}, \emph{1467}, 56--78\relax
\mciteBstWouldAddEndPuncttrue
\mciteSetBstMidEndSepPunct{\mcitedefaultmidpunct}
{\mcitedefaultendpunct}{\mcitedefaultseppunct}\relax
\EndOfBibitem
\bibitem[Weinberger(2000)]{WEINBERGER2000139}
Weinberger,~R. In \emph{Practical Capillary Electrophoresis (Second Edition)},
  second edition ed.; Weinberger,~R., Ed.; Academic Press: San Diego, 2000; pp
  139--208\relax
\mciteBstWouldAddEndPuncttrue
\mciteSetBstMidEndSepPunct{\mcitedefaultmidpunct}
{\mcitedefaultendpunct}{\mcitedefaultseppunct}\relax
\EndOfBibitem
\bibitem[Roberts and Caserio(1977)Roberts, and Caserio]{roberts1977basic}
Roberts,~J.~D.; Caserio,~M.~C. \emph{Basic principles of organic chemistry}; WA
  Benjamin, Inc., 1977\relax
\mciteBstWouldAddEndPuncttrue
\mciteSetBstMidEndSepPunct{\mcitedefaultmidpunct}
{\mcitedefaultendpunct}{\mcitedefaultseppunct}\relax
\EndOfBibitem
\bibitem[Govorov \latin{et~al.}(2010)Govorov, Fan, Hernandez, Slocik, and
  Naik]{Govorov2010}
Govorov,~A.~O.; Fan,~Z.; Hernandez,~P.; Slocik,~J.~M.; Naik,~R.~R. {Theory of
  circular dichroism of nanomaterials comprising chiral molecules and
  nanocrystals: Plasmon enhancement, dipole interactions, and dielectric
  effects}. \emph{Nano Letters} \textbf{2010}, \emph{10}, 1374--1382\relax
\mciteBstWouldAddEndPuncttrue
\mciteSetBstMidEndSepPunct{\mcitedefaultmidpunct}
{\mcitedefaultendpunct}{\mcitedefaultseppunct}\relax
\EndOfBibitem
\bibitem[Mauro \latin{et~al.}(2022)Mauro, Fregoni, Feist, and
  Avriller]{mauro2022chiral}
Mauro,~L.; Fregoni,~J.; Feist,~J.; Avriller,~R. Chiral Discrimination in
  Helicity-Preserving Fabry-P$\backslash$'erot Cavities. \emph{arXiv preprint
  arXiv:2209.00402} \textbf{2022}, \relax
\mciteBstWouldAddEndPunctfalse
\mciteSetBstMidEndSepPunct{\mcitedefaultmidpunct}
{}{\mcitedefaultseppunct}\relax
\EndOfBibitem
\bibitem[Riso \latin{et~al.}(2022)Riso, Grazioli, Ronca, Giovannini, and
  Koch]{riso2022strong}
Riso,~R.~R.; Grazioli,~L.; Ronca,~E.; Giovannini,~T.; Koch,~H. Strong coupling
  in chiral cavities: nonperturbative framework for enantiomer discrimination.
  \emph{arXiv preprint arXiv:2209.01987} \textbf{2022}, \relax
\mciteBstWouldAddEndPunctfalse
\mciteSetBstMidEndSepPunct{\mcitedefaultmidpunct}
{}{\mcitedefaultseppunct}\relax
\EndOfBibitem
\bibitem[Voronin \latin{et~al.}(2022)Voronin, Taradin, Gorkunov, and
  Baranov]{Voronin2022}
Voronin,~K.; Taradin,~A.~S.; Gorkunov,~M.~V.; Baranov,~D.~G. Single-handedness
  chiral optical cavities. \emph{ACS Photonics} \textbf{2022}, \emph{9},
  2652--2659\relax
\mciteBstWouldAddEndPuncttrue
\mciteSetBstMidEndSepPunct{\mcitedefaultmidpunct}
{\mcitedefaultendpunct}{\mcitedefaultseppunct}\relax
\EndOfBibitem
\bibitem[Sch{\"a}fer \latin{et~al.}(2021)Sch{\"a}fer, Buchholz, Penz,
  Ruggenthaler, and Rubio]{schafer2021making}
Sch{\"a}fer,~C.; Buchholz,~F.; Penz,~M.; Ruggenthaler,~M.; Rubio,~A. Making ab
  initio QED functional (s): Nonperturbative and photon-free effective
  frameworks for strong light--matter coupling. \emph{Proceedings of the
  National Academy of Sciences} \textbf{2021}, \emph{118}, e2110464118\relax
\mciteBstWouldAddEndPuncttrue
\mciteSetBstMidEndSepPunct{\mcitedefaultmidpunct}
{\mcitedefaultendpunct}{\mcitedefaultseppunct}\relax
\EndOfBibitem
\bibitem[Haugland \latin{et~al.}(2020)Haugland, Ronca, Kj{\o}nstad, Rubio, and
  Koch]{haugland2020coupled}
Haugland,~T.~S.; Ronca,~E.; Kj{\o}nstad,~E.~F.; Rubio,~A.; Koch,~H. Coupled
  cluster theory for molecular polaritons: Changing ground and excited states.
  \emph{Phys. Rev. X} \textbf{2020}, \emph{10}, 041043\relax
\mciteBstWouldAddEndPuncttrue
\mciteSetBstMidEndSepPunct{\mcitedefaultmidpunct}
{\mcitedefaultendpunct}{\mcitedefaultseppunct}\relax
\EndOfBibitem
\bibitem[Haugland \latin{et~al.}(2021)Haugland, Sch{\"a}fer, Ronca, Rubio, and
  Koch]{haugland2020intermolecular}
Haugland,~T.~S.; Sch{\"a}fer,~C.; Ronca,~E.; Rubio,~A.; Koch,~H. Intermolecular
  interactions in optical cavities: An ab initio QED study. \emph{J. Chem.
  Phys.} \textbf{2021}, \emph{154}, 094113\relax
\mciteBstWouldAddEndPuncttrue
\mciteSetBstMidEndSepPunct{\mcitedefaultmidpunct}
{\mcitedefaultendpunct}{\mcitedefaultseppunct}\relax
\EndOfBibitem
\bibitem[Flick \latin{et~al.}(2017)Flick, Ruggenthaler, Appel, and
  Rubio]{flick2017atoms}
Flick,~J.; Ruggenthaler,~M.; Appel,~H.; Rubio,~A. Atoms and molecules in
  cavities, from weak to strong coupling in quantum-electrodynamics (QED)
  chemistry. \emph{Proceedings of the National Academy of Sciences}
  \textbf{2017}, \emph{114}, 3026--3034\relax
\mciteBstWouldAddEndPuncttrue
\mciteSetBstMidEndSepPunct{\mcitedefaultmidpunct}
{\mcitedefaultendpunct}{\mcitedefaultseppunct}\relax
\EndOfBibitem
\bibitem[Ruggenthaler \latin{et~al.}(2018)Ruggenthaler, Tancogne-Dejean, Flick,
  Appel, and Rubio]{ruggenthaler2018quantum}
Ruggenthaler,~M.; Tancogne-Dejean,~N.; Flick,~J.; Appel,~H.; Rubio,~A. From a
  quantum-electrodynamical light--matter description to novel spectroscopies.
  \emph{Nat. Rev. Chem.} \textbf{2018}, \emph{2}, 0118\relax
\mciteBstWouldAddEndPuncttrue
\mciteSetBstMidEndSepPunct{\mcitedefaultmidpunct}
{\mcitedefaultendpunct}{\mcitedefaultseppunct}\relax
\EndOfBibitem
\bibitem[Mason(2013)]{mason2013optical}
Mason,~S.~F. \emph{Optical activity and chiral discrimination}; Springer
  Science \& Business Media, 2013; Vol.~48\relax
\mciteBstWouldAddEndPuncttrue
\mciteSetBstMidEndSepPunct{\mcitedefaultmidpunct}
{\mcitedefaultendpunct}{\mcitedefaultseppunct}\relax
\EndOfBibitem
\bibitem[Li and Shapiro(2010)Li, and Shapiro]{li2010theory}
Li,~X.; Shapiro,~M. Theory of the optical spatial separation of racemic
  mixtures of chiral molecules. \emph{The Journal of chemical physics}
  \textbf{2010}, \emph{132}, 194315\relax
\mciteBstWouldAddEndPuncttrue
\mciteSetBstMidEndSepPunct{\mcitedefaultmidpunct}
{\mcitedefaultendpunct}{\mcitedefaultseppunct}\relax
\EndOfBibitem
\bibitem[Forbes and Andrews(2021)Forbes, and Andrews]{forbes2021orbital}
Forbes,~K.~A.; Andrews,~D.~L. Orbital angular momentum of twisted light:
  chirality and optical activity. \emph{Journal of Physics: Photonics}
  \textbf{2021}, \emph{3}, 022007\relax
\mciteBstWouldAddEndPuncttrue
\mciteSetBstMidEndSepPunct{\mcitedefaultmidpunct}
{\mcitedefaultendpunct}{\mcitedefaultseppunct}\relax
\EndOfBibitem
\bibitem[Babiker \latin{et~al.}(1974)Babiker, Power, and
  Thirunamachandran]{babiker1974generalization}
Babiker,~M.; Power,~E.~A.; Thirunamachandran,~T. On a generalization of the
  Power--Zienau--Woolley transformation in quantum electrodynamics and atomic
  field equations. \emph{Proceedings of the Royal Society of London. A.
  Mathematical and Physical Sciences} \textbf{1974}, \emph{338}, 235--249\relax
\mciteBstWouldAddEndPuncttrue
\mciteSetBstMidEndSepPunct{\mcitedefaultmidpunct}
{\mcitedefaultendpunct}{\mcitedefaultseppunct}\relax
\EndOfBibitem
\bibitem[Craig and Manolopoulos(2004)Craig, and Manolopoulos]{craig2004quantum}
Craig,~I.~R.; Manolopoulos,~D.~E. Quantum statistics and classical mechanics:
  Real time correlation functions from ring polymer molecular dynamics.
  \emph{J. Chem. Phys.} \textbf{2004}, \emph{121}, 3368--3373\relax
\mciteBstWouldAddEndPuncttrue
\mciteSetBstMidEndSepPunct{\mcitedefaultmidpunct}
{\mcitedefaultendpunct}{\mcitedefaultseppunct}\relax
\EndOfBibitem
\bibitem[Forbes(2018)]{forbes2018}
Forbes,~K.~A. Role of magnetic and diamagnetic interactions in molecular optics
  and scattering. \emph{Phys. Rev. A} \textbf{2018}, \emph{97}, 053832\relax
\mciteBstWouldAddEndPuncttrue
\mciteSetBstMidEndSepPunct{\mcitedefaultmidpunct}
{\mcitedefaultendpunct}{\mcitedefaultseppunct}\relax
\EndOfBibitem
\bibitem[Andrews \latin{et~al.}(2018)Andrews, Jones, Salam, and
  Woolley]{andrews2018perspective}
Andrews,~D.~L.; Jones,~G.~A.; Salam,~A.; Woolley,~R.~G. Perspective: Quantum
  Hamiltonians for optical interactions. \emph{The Journal of chemical physics}
  \textbf{2018}, \emph{148}, 040901\relax
\mciteBstWouldAddEndPuncttrue
\mciteSetBstMidEndSepPunct{\mcitedefaultmidpunct}
{\mcitedefaultendpunct}{\mcitedefaultseppunct}\relax
\EndOfBibitem
\bibitem[Sch\"afer \latin{et~al.}(2020)Sch\"afer, Ruggenthaler, Rokaj, and
  Rubio]{schafer2019relevance}
Sch\"afer,~C.; Ruggenthaler,~M.; Rokaj,~V.; Rubio,~A. Relevance of the
  quadratic diamagnetic and self-polarization terms in cavity quantum
  electrodynamics. \emph{ACS Photonics} \textbf{2020}, \emph{7}, 975--990\relax
\mciteBstWouldAddEndPuncttrue
\mciteSetBstMidEndSepPunct{\mcitedefaultmidpunct}
{\mcitedefaultendpunct}{\mcitedefaultseppunct}\relax
\EndOfBibitem
\bibitem[Power and Thirunamachandran(1982)Power, and
  Thirunamachandran]{power1982quantum}
Power,~E.~A.; Thirunamachandran,~T. Quantum electrodynamics in a cavity.
  \emph{Phys. Rev. A} \textbf{1982}, \emph{25}, 2473\relax
\mciteBstWouldAddEndPuncttrue
\mciteSetBstMidEndSepPunct{\mcitedefaultmidpunct}
{\mcitedefaultendpunct}{\mcitedefaultseppunct}\relax
\EndOfBibitem
\bibitem[Viehmann \latin{et~al.}(2011)Viehmann, von Delft, and
  Marquardt]{viehmann2011superradiant}
Viehmann,~O.; von Delft,~J.; Marquardt,~F. Superradiant phase transitions and
  the standard description of circuit QED. \emph{Phys. Rev. Lett.}
  \textbf{2011}, \emph{107}, 113602\relax
\mciteBstWouldAddEndPuncttrue
\mciteSetBstMidEndSepPunct{\mcitedefaultmidpunct}
{\mcitedefaultendpunct}{\mcitedefaultseppunct}\relax
\EndOfBibitem
\bibitem[Ashida \latin{et~al.}(2020)Ashida, \ifmmode \dot{I}\else
  \.{I}\fi{}mamo\ifmmode~\breve{g}\else \u{g}\fi{}lu, Faist, Jaksch, Cavalleri,
  and Demler]{ashida2020}
Ashida,~Y.; \ifmmode \dot{I}\else \.{I}\fi{}mamo\ifmmode~\breve{g}\else
  \u{g}\fi{}lu,~A. m.~c.; Faist,~J.; Jaksch,~D.; Cavalleri,~A.; Demler,~E.
  Quantum Electrodynamic Control of Matter: Cavity-Enhanced Ferroelectric Phase
  Transition. \emph{Phys. Rev. X} \textbf{2020}, \emph{10}, 041027\relax
\mciteBstWouldAddEndPuncttrue
\mciteSetBstMidEndSepPunct{\mcitedefaultmidpunct}
{\mcitedefaultendpunct}{\mcitedefaultseppunct}\relax
\EndOfBibitem
\bibitem[Lenk \latin{et~al.}(2022)Lenk, Li, Werner, and Eckstein]{lenk2022}
Lenk,~K.; Li,~J.; Werner,~P.; Eckstein,~M. Collective theory for an interacting
  solid in a single-mode cavity. \emph{arXiv preprint arXiv:2205.05559}
  \textbf{2022}, \relax
\mciteBstWouldAddEndPunctfalse
\mciteSetBstMidEndSepPunct{\mcitedefaultmidpunct}
{}{\mcitedefaultseppunct}\relax
\EndOfBibitem
\bibitem[De~Bernardis \latin{et~al.}(2018)De~Bernardis, Jaako, and
  Rabl]{de2018cavity}
De~Bernardis,~D.; Jaako,~T.; Rabl,~P. Cavity quantum electrodynamics in the
  nonperturbative regime. \emph{Phys. Rev. A} \textbf{2018}, \emph{97},
  043820\relax
\mciteBstWouldAddEndPuncttrue
\mciteSetBstMidEndSepPunct{\mcitedefaultmidpunct}
{\mcitedefaultendpunct}{\mcitedefaultseppunct}\relax
\EndOfBibitem
\bibitem[Nataf \latin{et~al.}(2019)Nataf, Champel, Blatter, and
  Basko]{nataf2019}
Nataf,~P.; Champel,~T.; Blatter,~G.; Basko,~D.~M. Rashba Cavity QED: A Route
  Towards the Superradiant Quantum Phase Transition. \emph{Phys. Rev. Lett.}
  \textbf{2019}, \emph{123}, 207402\relax
\mciteBstWouldAddEndPuncttrue
\mciteSetBstMidEndSepPunct{\mcitedefaultmidpunct}
{\mcitedefaultendpunct}{\mcitedefaultseppunct}\relax
\EndOfBibitem
\bibitem[Andolina \latin{et~al.}(2020)Andolina, Pellegrino, Giovannetti,
  MacDonald, and Polini]{andolina2020}
Andolina,~G.~M.; Pellegrino,~F. M.~D.; Giovannetti,~V.; MacDonald,~A.~H.;
  Polini,~M. Theory of photon condensation in a spatially varying
  electromagnetic field. \emph{Phys. Rev. B} \textbf{2020}, \emph{102},
  125137\relax
\mciteBstWouldAddEndPuncttrue
\mciteSetBstMidEndSepPunct{\mcitedefaultmidpunct}
{\mcitedefaultendpunct}{\mcitedefaultseppunct}\relax
\EndOfBibitem
\bibitem[Baranov \latin{et~al.}(2020)Baranov, Munkhbat, L{\"{a}}nk, Verre,
  K{\"{a}}ll, and Shegai]{Baranov2020CD}
Baranov,~D.~G.; Munkhbat,~B.; L{\"{a}}nk,~N.~O.; Verre,~R.; K{\"{a}}ll,~M.;
  Shegai,~T. {Circular dichroism mode splitting and bounds to its enhancement
  with cavity-plasmon-polaritons}. \emph{Nanophotonics} \textbf{2020},
  \emph{9}, 283--293\relax
\mciteBstWouldAddEndPuncttrue
\mciteSetBstMidEndSepPunct{\mcitedefaultmidpunct}
{\mcitedefaultendpunct}{\mcitedefaultseppunct}\relax
\EndOfBibitem
\bibitem[Semnani \latin{et~al.}(2020)Semnani, Flannery, {Al Maruf}, and
  Bajcsy]{Semnani2020}
Semnani,~B.; Flannery,~J.; {Al Maruf},~R.; Bajcsy,~M. {Spin-preserving chiral
  photonic crystal mirror}. \emph{Light: Science and Applications}
  \textbf{2020}, \emph{9}, 23\relax
\mciteBstWouldAddEndPuncttrue
\mciteSetBstMidEndSepPunct{\mcitedefaultmidpunct}
{\mcitedefaultendpunct}{\mcitedefaultseppunct}\relax
\EndOfBibitem
\bibitem[Fernandez-Corbaton \latin{et~al.}(2016)Fernandez-Corbaton, Fruhnert,
  and Rockstuhl]{Fernandez-Corbaton2016}
Fernandez-Corbaton,~I.; Fruhnert,~M.; Rockstuhl,~C. {Objects of maximum
  electromagnetic chirality}. \emph{Physical Review X} \textbf{2016}, \emph{6},
  031013\relax
\mciteBstWouldAddEndPuncttrue
\mciteSetBstMidEndSepPunct{\mcitedefaultmidpunct}
{\mcitedefaultendpunct}{\mcitedefaultseppunct}\relax
\EndOfBibitem
\bibitem[Sch\"afer \latin{et~al.}(2018)Sch\"afer, Ruggenthaler, and
  Rubio]{schafer2018insights}
Sch\"afer,~C.; Ruggenthaler,~M.; Rubio,~A. Ab initio nonrelativistic quantum
  electrodynamics: Bridging quantum chemistry and quantum optics from weak to
  strong coupling. \emph{Phys. Rev. A} \textbf{2018}, \emph{98}, 043801\relax
\mciteBstWouldAddEndPuncttrue
\mciteSetBstMidEndSepPunct{\mcitedefaultmidpunct}
{\mcitedefaultendpunct}{\mcitedefaultseppunct}\relax
\EndOfBibitem
\bibitem[Hopfield(1958)]{hopfield1958}
Hopfield,~J.~J. Theory of the Contribution of Excitons to the Complex
  Dielectric Constant of Crystals. \emph{Phys. Rev.} \textbf{1958}, \emph{112},
  1555--1567\relax
\mciteBstWouldAddEndPuncttrue
\mciteSetBstMidEndSepPunct{\mcitedefaultmidpunct}
{\mcitedefaultendpunct}{\mcitedefaultseppunct}\relax
\EndOfBibitem
\bibitem[George \latin{et~al.}(2016)George, Chervy, Shalabney, Devaux, Hiura,
  Genet, and Ebbesen]{george2016}
George,~J.; Chervy,~T.; Shalabney,~A.; Devaux,~E.; Hiura,~H.; Genet,~C.;
  Ebbesen,~T.~W. Multiple Rabi Splittings under Ultrastrong Vibrational
  Coupling. \emph{Phys. Rev. Lett.} \textbf{2016}, \emph{117}, 153601\relax
\mciteBstWouldAddEndPuncttrue
\mciteSetBstMidEndSepPunct{\mcitedefaultmidpunct}
{\mcitedefaultendpunct}{\mcitedefaultseppunct}\relax
\EndOfBibitem
\bibitem[Todorov and Sirtori(2012)Todorov, and Sirtori]{todorov2012}
Todorov,~Y.; Sirtori,~C. Intersubband polaritons in the electrical dipole
  gauge. \emph{Phys. Rev. B} \textbf{2012}, \emph{85}, 045304\relax
\mciteBstWouldAddEndPuncttrue
\mciteSetBstMidEndSepPunct{\mcitedefaultmidpunct}
{\mcitedefaultendpunct}{\mcitedefaultseppunct}\relax
\EndOfBibitem
\bibitem[doi(2005)]{doi:10.1126/science.309.5731.78b}
So Much More to Know. \emph{Science} \textbf{2005}, \emph{309}, 78--102\relax
\mciteBstWouldAddEndPuncttrue
\mciteSetBstMidEndSepPunct{\mcitedefaultmidpunct}
{\mcitedefaultendpunct}{\mcitedefaultseppunct}\relax
\EndOfBibitem
\bibitem[Blackmond(2004)]{blackmond2004}
Blackmond,~D.~G. Asymmetric autocatalysis and its implications for the origin
  of homochirality. \emph{Proceedings of the National Academy of Sciences}
  \textbf{2004}, \emph{101}, 5732--5736\relax
\mciteBstWouldAddEndPuncttrue
\mciteSetBstMidEndSepPunct{\mcitedefaultmidpunct}
{\mcitedefaultendpunct}{\mcitedefaultseppunct}\relax
\EndOfBibitem
\bibitem[Hertzog \latin{et~al.}(2021)Hertzog, Munkhbat, Baranov, Shegai, and
  B\"orjesson]{hertzog2021enhancing}
Hertzog,~M.; Munkhbat,~B.; Baranov,~D.; Shegai,~T.; B\"orjesson,~K. Enhancing
  Vibrational Light--Matter Coupling Strength beyond the Molecular
  Concentration Limit Using Plasmonic Arrays. \emph{Nano letters}
  \textbf{2021}, \emph{21}, 1320--1326\relax
\mciteBstWouldAddEndPuncttrue
\mciteSetBstMidEndSepPunct{\mcitedefaultmidpunct}
{\mcitedefaultendpunct}{\mcitedefaultseppunct}\relax
\EndOfBibitem
\end{mcitethebibliography}


\begin{thebibliography}{11}%
\makeatletter
\providecommand \@ifxundefined [1]{%
 \@ifx{#1\undefined}
}%
\providecommand \@ifnum [1]{%
 \ifnum #1\expandafter \@firstoftwo
 \else \expandafter \@secondoftwo
 \fi
}%
\providecommand \@ifx [1]{%
 \ifx #1\expandafter \@firstoftwo
 \else \expandafter \@secondoftwo
 \fi
}%
\providecommand \natexlab [1]{#1}%
\providecommand \enquote  [1]{``#1''}%
\providecommand \bibnamefont  [1]{#1}%
\providecommand \bibfnamefont [1]{#1}%
\providecommand \citenamefont [1]{#1}%
\providecommand \href@noop [0]{\@secondoftwo}%
\providecommand \href [0]{\begingroup \@sanitize@url \@href}%
\providecommand \@href[1]{\@@startlink{#1}\@@href}%
\providecommand \@@href[1]{\endgroup#1\@@endlink}%
\providecommand \@sanitize@url [0]{\catcode `\\12\catcode `\$12\catcode
  `\&12\catcode `\#12\catcode `\^12\catcode `\_12\catcode `\%12\relax}%
\providecommand \@@startlink[1]{}%
\providecommand \@@endlink[0]{}%
\providecommand \url  [0]{\begingroup\@sanitize@url \@url }%
\providecommand \@url [1]{\endgroup\@href {#1}{\urlprefix }}%
\providecommand \urlprefix  [0]{URL }%
\providecommand \Eprint [0]{\href }%
\providecommand \doibase [0]{http://dx.doi.org/}%
\providecommand \selectlanguage [0]{\@gobble}%
\providecommand \bibinfo  [0]{\@secondoftwo}%
\providecommand \bibfield  [0]{\@secondoftwo}%
\providecommand \translation [1]{[#1]}%
\providecommand \BibitemOpen [0]{}%
\providecommand \bibitemStop [0]{}%
\providecommand \bibitemNoStop [0]{.\EOS\space}%
\providecommand \EOS [0]{\spacefactor3000\relax}%
\providecommand \BibitemShut  [1]{\csname bibitem#1\endcsname}%
\let\auto@bib@innerbib\@empty
\bibitem [{\citenamefont {Sch\"afer}\ \emph {et~al.}(2020)\citenamefont
  {Sch\"afer}, \citenamefont {Ruggenthaler}, \citenamefont {Rokaj},\ and\
  \citenamefont {Rubio}}]{schafer2019relevance}%
  \BibitemOpen
  \bibfield  {author} {\bibinfo {author} {\bibfnamefont {C.}~\bibnamefont
  {Sch\"afer}}, \bibinfo {author} {\bibfnamefont {M.}~\bibnamefont
  {Ruggenthaler}}, \bibinfo {author} {\bibfnamefont {V.}~\bibnamefont {Rokaj}},
  \ and\ \bibinfo {author} {\bibfnamefont {A.}~\bibnamefont {Rubio}},\ }\href
  {\doibase 10.1021/acsphotonics.9b01649} {\bibfield  {journal} {\bibinfo
  {journal} {ACS Photonics}\ }\textbf {\bibinfo {volume} {7}},\ \bibinfo
  {pages} {975} (\bibinfo {year} {2020})}\BibitemShut {NoStop}%
\bibitem [{\citenamefont {Rokaj}\ \emph {et~al.}(2018)\citenamefont {Rokaj},
  \citenamefont {Welakuh}, \citenamefont {Ruggenthaler},\ and\ \citenamefont
  {Rubio}}]{rokaj2017}%
  \BibitemOpen
  \bibfield  {author} {\bibinfo {author} {\bibfnamefont {V.}~\bibnamefont
  {Rokaj}}, \bibinfo {author} {\bibfnamefont {D.~M.}\ \bibnamefont {Welakuh}},
  \bibinfo {author} {\bibfnamefont {M.}~\bibnamefont {Ruggenthaler}}, \ and\
  \bibinfo {author} {\bibfnamefont {A.}~\bibnamefont {Rubio}},\ }\href
  {\doibase 10.1088/1361-6455/aa9c99} {\bibfield  {journal} {\bibinfo
  {journal} {J. Phys. B}\ }\textbf {\bibinfo {volume} {51}},\ \bibinfo {pages}
  {034005} (\bibinfo {year} {2018})}\BibitemShut {NoStop}%
\bibitem [{\citenamefont {Condon}(1937)}]{Condon1937}%
  \BibitemOpen
  \bibfield  {author} {\bibinfo {author} {\bibfnamefont {E.~U.}\ \bibnamefont
  {Condon}},\ }\href {\doibase 10.1103/RevModPhys.9.432} {\bibfield  {journal}
  {\bibinfo  {journal} {Reviews of Modern Physics}\ }\textbf {\bibinfo {volume}
  {9}},\ \bibinfo {pages} {432} (\bibinfo {year} {1937})}\BibitemShut {NoStop}%
\bibitem [{\citenamefont {Corbaton}(2014)}]{corbaton2014helicity}%
  \BibitemOpen
  \bibfield  {author} {\bibinfo {author} {\bibfnamefont {I.~F.}\ \bibnamefont
  {Corbaton}},\ }\emph {\bibinfo {title} {Helicity and duality symmetry in
  light matter interactions: Theory and applications}},\ \href@noop {} {Ph.D.
  thesis},\ \bibinfo  {school} {Macquarie University, Faculty of Science and
  Engineering} (\bibinfo {year} {2014})\BibitemShut {NoStop}%
\bibitem [{\citenamefont {Caloz}\ \emph {et~al.}(2018)\citenamefont {Caloz},
  \citenamefont {Alu}, \citenamefont {Tretyakov}, \citenamefont {Sounas},
  \citenamefont {Achouri},\ and\ \citenamefont
  {Deck-L{\'e}ger}}]{caloz2018electromagnetic}%
  \BibitemOpen
  \bibfield  {author} {\bibinfo {author} {\bibfnamefont {C.}~\bibnamefont
  {Caloz}}, \bibinfo {author} {\bibfnamefont {A.}~\bibnamefont {Alu}}, \bibinfo
  {author} {\bibfnamefont {S.}~\bibnamefont {Tretyakov}}, \bibinfo {author}
  {\bibfnamefont {D.}~\bibnamefont {Sounas}}, \bibinfo {author} {\bibfnamefont
  {K.}~\bibnamefont {Achouri}}, \ and\ \bibinfo {author} {\bibfnamefont
  {Z.-L.}\ \bibnamefont {Deck-L{\'e}ger}},\ }\href@noop {} {\bibfield
  {journal} {\bibinfo  {journal} {Physical Review Applied}\ }\textbf {\bibinfo
  {volume} {10}},\ \bibinfo {pages} {047001} (\bibinfo {year}
  {2018})}\BibitemShut {NoStop}%
\bibitem [{\citenamefont {Hopfield}(1958)}]{hopfield1958}%
  \BibitemOpen
  \bibfield  {author} {\bibinfo {author} {\bibfnamefont {J.~J.}\ \bibnamefont
  {Hopfield}},\ }\href {\doibase 10.1103/PhysRev.112.1555} {\bibfield
  {journal} {\bibinfo  {journal} {Phys. Rev.}\ }\textbf {\bibinfo {volume}
  {112}},\ \bibinfo {pages} {1555} (\bibinfo {year} {1958})}\BibitemShut
  {NoStop}%
\bibitem [{\citenamefont {Todorov}\ and\ \citenamefont
  {Sirtori}(2012)}]{todorov2012}%
  \BibitemOpen
  \bibfield  {author} {\bibinfo {author} {\bibfnamefont {Y.}~\bibnamefont
  {Todorov}}\ and\ \bibinfo {author} {\bibfnamefont {C.}~\bibnamefont
  {Sirtori}},\ }\href {\doibase 10.1103/PhysRevB.85.045304} {\bibfield
  {journal} {\bibinfo  {journal} {Phys. Rev. B}\ }\textbf {\bibinfo {volume}
  {85}},\ \bibinfo {pages} {045304} (\bibinfo {year} {2012})}\BibitemShut
  {NoStop}%
\bibitem [{\citenamefont {Voronin}\ \emph {et~al.}(2022)\citenamefont
  {Voronin}, \citenamefont {Taradin}, \citenamefont {Gorkunov},\ and\
  \citenamefont {Baranov}}]{Voronin2022}%
  \BibitemOpen
  \bibfield  {author} {\bibinfo {author} {\bibfnamefont {K.}~\bibnamefont
  {Voronin}}, \bibinfo {author} {\bibfnamefont {A.~S.}\ \bibnamefont
  {Taradin}}, \bibinfo {author} {\bibfnamefont {M.~V.}\ \bibnamefont
  {Gorkunov}}, \ and\ \bibinfo {author} {\bibfnamefont {D.~G.}\ \bibnamefont
  {Baranov}},\ }\href@noop {} {\bibfield  {journal} {\bibinfo  {journal} {ACS
  Photonics}\ }\textbf {\bibinfo {volume} {9}},\ \bibinfo {pages} {2652}
  (\bibinfo {year} {2022})}\BibitemShut {NoStop}%
\bibitem [{\citenamefont {Tichauer}\ \emph {et~al.}(2021)\citenamefont
  {Tichauer}, \citenamefont {Feist},\ and\ \citenamefont
  {Groenhof}}]{tichauer2021multi}%
  \BibitemOpen
  \bibfield  {author} {\bibinfo {author} {\bibfnamefont {R.~H.}\ \bibnamefont
  {Tichauer}}, \bibinfo {author} {\bibfnamefont {J.}~\bibnamefont {Feist}}, \
  and\ \bibinfo {author} {\bibfnamefont {G.}~\bibnamefont {Groenhof}},\
  }\href@noop {} {\bibfield  {journal} {\bibinfo  {journal} {The Journal of
  Chemical Physics}\ }\textbf {\bibinfo {volume} {154}},\ \bibinfo {pages}
  {104112} (\bibinfo {year} {2021})}\BibitemShut {NoStop}%
\bibitem [{\citenamefont {George}\ \emph {et~al.}(2016)\citenamefont {George},
  \citenamefont {Chervy}, \citenamefont {Shalabney}, \citenamefont {Devaux},
  \citenamefont {Hiura}, \citenamefont {Genet},\ and\ \citenamefont
  {Ebbesen}}]{george2016}%
  \BibitemOpen
  \bibfield  {author} {\bibinfo {author} {\bibfnamefont {J.}~\bibnamefont
  {George}}, \bibinfo {author} {\bibfnamefont {T.}~\bibnamefont {Chervy}},
  \bibinfo {author} {\bibfnamefont {A.}~\bibnamefont {Shalabney}}, \bibinfo
  {author} {\bibfnamefont {E.}~\bibnamefont {Devaux}}, \bibinfo {author}
  {\bibfnamefont {H.}~\bibnamefont {Hiura}}, \bibinfo {author} {\bibfnamefont
  {C.}~\bibnamefont {Genet}}, \ and\ \bibinfo {author} {\bibfnamefont {T.~W.}\
  \bibnamefont {Ebbesen}},\ }\href {\doibase 10.1103/PhysRevLett.117.153601}
  {\bibfield  {journal} {\bibinfo  {journal} {Phys. Rev. Lett.}\ }\textbf
  {\bibinfo {volume} {117}},\ \bibinfo {pages} {153601} (\bibinfo {year}
  {2016})}\BibitemShut {NoStop}%
\bibitem [{\citenamefont {Riso}\ \emph {et~al.}(2022)\citenamefont {Riso},
  \citenamefont {Grazioli}, \citenamefont {Ronca}, \citenamefont {Giovannini},\
  and\ \citenamefont {Koch}}]{riso2022strong}%
  \BibitemOpen
  \bibfield  {author} {\bibinfo {author} {\bibfnamefont {R.~R.}\ \bibnamefont
  {Riso}}, \bibinfo {author} {\bibfnamefont {L.}~\bibnamefont {Grazioli}},
  \bibinfo {author} {\bibfnamefont {E.}~\bibnamefont {Ronca}}, \bibinfo
  {author} {\bibfnamefont {T.}~\bibnamefont {Giovannini}}, \ and\ \bibinfo
  {author} {\bibfnamefont {H.}~\bibnamefont {Koch}},\ }\href@noop {} {\bibfield
   {journal} {\bibinfo  {journal} {arXiv preprint arXiv:2209.01987}\ }
  (\bibinfo {year} {2022})}\BibitemShut {NoStop}%
\end{thebibliography}%

\end{document}


\title{Supplemental Information to Chiral Polaritonics: Analytic Solutions, Intuition and its Use}

\author{Christian Sch\"afer}
\email[]{christian.schaefer.physics@gmail.com}
\affiliation{MC2 Department, Chalmers University of Technology, Sweden}

\author{Denis G. Baranov}
\email[]{denis.baranov@phystech.edu}
\affiliation{Center for Photonics and 2D Materials, Moscow Institute of Physics and Technology, Dolgoprudny 141700, Russia}

\begin{abstract}
The Supplemental Information provides an extended derivation of the chiral Hopfield and Tavis-Cummings model and an alternative derivation of the chiral Hopfield model in which the self-polarization terms are partially cancelled which results in an instability of the chiral system.
\end{abstract}

\maketitle

\section{Fields of a standing chiral wave}

\rev{
In the following we obtain explicit expressions for the electric and magnetic fields of a chiral standing wave describing the optical mode of a single-handedness optical cavity. We begin with the case of a standing wave formed by two counter-propagating circularly polarized plane waves. 
The field of a monochromatic circularly polarized plane wave propagating through air in the positive direction of the z axis takes the form ($e^{-i \omega t}$ time dependence for the harmonic field is assumed):
\begin{equation}
    \bE^{\lambda}_{+z}(\br) = \frac{\mE}{\sqrt{2}} 
    \begin{pmatrix}
    1 \\
    i \lambda \\
    0
    \end{pmatrix} e^{i k z},\ \ 
    Z\bH^{\lambda}_{+z} (\br) = - i \lambda \bE^{\lambda}_{+z}(\br),
\end{equation}
where $\lambda = \pm 1$ denotes the handedness of the wave, $k = \omega / c$, and $\mE$ has units of electric field. Correspondingly, the fields of a wave travelling in the negative direction of the z axis takes the form:
\begin{equation}
    \bE^{\lambda}_{-z}(\br) = \frac{\mE}{\sqrt{2}} 
    \begin{pmatrix}
    1 \\
    -i \lambda \\
    0
    \end{pmatrix} e^{-i k z},\ \ 
    Z\bH^{\lambda}_{-z} (\br) = - i \lambda \bE^{\lambda}_{-z}(\br).
\end{equation}
The fields of a "vertical" standing wave take the form:
\begin{equation}
    \bE^{\lambda}_{\bk_{\parallel} = 0} = \frac{\bE^{\lambda}_{+z} + \bE^{\lambda}_{-z}}{ \sqrt{2} } = 
    \mE
    \begin{pmatrix}
    \cos kz \\
    - \lambda \sin kz \\
    0
    \end{pmatrix},\ \ 
    Z \bH^{\lambda}_{\bk_{\parallel} = 0} =
    - i \lambda \bE^{\lambda}_{\bk_{\parallel} = 0}(\br).
\end{equation}
}

\rev{
Now consider a pair of circularly polarized plane waves with a given handedness $\lambda$ both propagating with a fixed in-plane momentum $\bk_{\parallel}$ (in-plane with respect to the vertical axis of the cavity) and opposite vertical component of the wave vector $\pm k_z \uz$. 
Without loss of generality let us assume $\bk_{\parallel} = k_x \ux$ with $k_x = k \sin \theta$ and $k_z = \pm k \cos \theta$. 
Electric fields take the form:
\begin{equation}
    \bE^{\lambda}_{+z,\bk_{\parallel}}(\br) = \frac{\mE}{\sqrt{2}} 
    \begin{pmatrix}
    \cos \theta \\
    i \lambda \\
    - \sin \theta
    \end{pmatrix} e^{i k_z z + i k_x x},\ \ 
    \bE^{\lambda}_{-z,\bk_{\parallel}}(\br) = \frac{\mE}{\sqrt{2}} 
    \begin{pmatrix}
    \cos \theta \\
    - i \lambda \\
    + \sin \theta
    \end{pmatrix} e^{-i k_z z + i k_x x}
\end{equation}
The z-dependent electric field of the combination of the two waves with the common $\bk_{\parallel}$ takes the form:
\begin{equation}
    \bE^{\lambda}_{\bk_{\parallel}}(\br) = \mE
    \begin{pmatrix}
    \cos \theta \cos k_z z \\
    - \lambda \sin k_z z \\
    - i \sin \theta \sin k_z z
    \end{pmatrix} e^{i k_x x}.
\end{equation}
Magnetic field of the chiral standing wave with handedness $\lambda$ follows from the electric field:
\begin{equation}
     Z \bH^{\lambda}_{\bk_{\parallel}} = - i \lambda   \bE^{\lambda}_{\bk_{\parallel}}.
\end{equation}
}

\begin{figure}[h!]
\includegraphics[width=.9\textwidth]{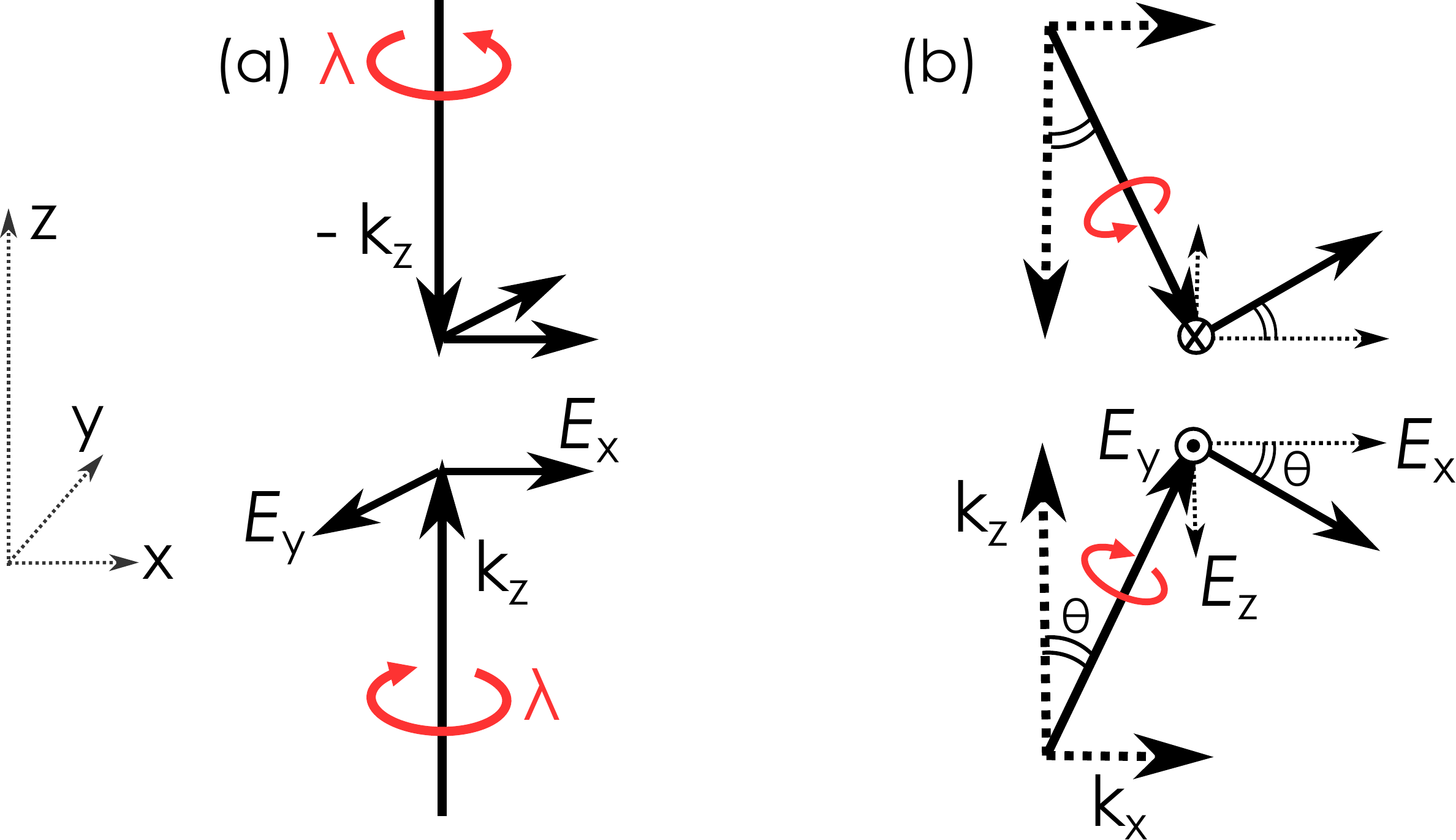}
\caption{Geometry of a chiral standing wave.}
\label{fig_S1}
\end{figure}

\newpage

\section{Deriving the chiral Hopfield and Tavis-Cummings models}
We start with the chiral standing fields that are derived from the generic mode expansion
\begin{align*}
    \hat{\bD}_\perp(\br) &= i \sum_{\bk,\lambda} \sqrt{\frac{\hbar c k \eps_0}{2 V}} \big( \epsilon_{\bk\lambda} e^{i\bk \cdot \br} \ha _{\bk\lambda} - \epsilon_{\bk\lambda}^* e^{-i\bk \cdot \br} \ha ^\dagger_{\bk\lambda} \big),\\
    \hat{\bB}(\br) &= i \frac{1}{c} \sum_{\bk,\lambda} \sqrt{\frac{\hbar c k }{2 \eps_0 V}} \big( \beta_{\bk\lambda} e^{i\bk \cdot \br} \ha _{\bk\lambda} - \beta_{\bk\lambda}^* e^{-i\bk \cdot \br} \ha ^\dagger_{\bk\lambda} \big).
\end{align*}
Following the explanations in the main text, 'upwards' and 'downwards' propagating fields are superimposed as
\begin{align*}
    \hat{\bD}^{\lambda}_\perp(\br) &= \sum_{\bk, \lambda} \hat{\bD}^{\lambda}_{\bk,\perp}(\br), \\
    \hat{\bD}^{\lambda}_\perp(\br) &= \sum_k (\hat{\bD}^{\lambda}_{+k,\perp}(\br) + \hat{\bD}^{\lambda}_{-k,\perp}(\br) )/\sqrt{2},
\end{align*}
with $\hat{a}_{+k_z,\lambda} = \hat{a}_{-k_z,\lambda}$ and $ \epsilon_{\pm k,\lambda}^{} = \frac{1}{\sqrt{2}}(1, \pm i \lambda, 0)^T$.
One obtains 
\begin{align*}
    \hat{\bD}^{\lambda}_\perp(\br) = - \sum_{k>0} \sqrt{\frac{\eps_0}{V}} \tilde{\eps}^{\lambda}_{k}(z) \hat{p}_{k,\lambda},
\end{align*}
with $\tilde{\eps}^{\lambda}_{k}(z) =  \left(\cos(k z), - \lambda \sin(k z), 0 \right)^T$ and $\hat{p}_{k,\lambda} = -i\sqrt{\hbar c k/2} ( \ha _{\bk,\lambda} - \ha ^\dagger_{\bk,\lambda}) $, $\hat{q}_{k,\lambda} = \sqrt{\hbar/2ck} ( \ha _{\bk,\lambda} + \ha ^\dagger_{\bk,\lambda} )$. The same procedure is applied to the magnetic fields where now however $\beta_{\bk,\lambda} = -i \lambda \epsilon_{\bk,\lambda}$ such that the chiral standing magnetic field features the additional $\lambda$
\begin{align*}
    \hat{\bB}_{\lambda}(\br) = \sum_{k > 0} \sqrt{k^2/\eps_0 V} \lambda \tilde{\eps}^{\lambda}_{k}(z) \hat{q}_{k,\lambda}.
\end{align*}
Considering that $\nabla \times \tilde{\eps}^{\lambda}_{k}(z) = \lambda k \tilde{\eps}^{\lambda}_{k}(z)$, it is easily validated that those fields fulfill Maxwell's equations of motion and contribute the common photonic energy $\sum_{k>0} \hbar \omega_k (\hat{a}^\dagger \hat{a}+\frac{1}{2})$ per handedness $\lambda$.

We will assume in the following that only a single mode couples substantially to the ensemble of chiral molecules, a reasonable approximation for most cavity realizations that should be however relaxed if the inter-molecular distances become much larger than the wavelength of the electromagnetic standing field.
Explicitly expressing all components in the Hamiltonian leads to
\begin{equation}
\begin{split}
    \hat{H} &= \sum_n \hat{H}_{M,n} + \frac{1}{2\varepsilon_0 V}\Big[\sum_n \tilde{\eps}^{\lambda}_{k}(z) \cdot \hat{\bmu }_n\Big]^2
    + \frac{1}{2}\hat{p}_{k}^2 \\
    &+ \frac{1}{2}\big[ \omega_{k}^2 + \sum_{n=1}^{N_{mol}} \sum_{i,j=1}^{3} 2\hat{\chi}^m_{n,ij} \frac{k^2}{\varepsilon_0 V} \tilde{\eps}^{\lambda}_{k,i}(z) \tilde{\eps}^{\lambda}_{k,j}(z) \big] \hat{q}_{k}^2\\
    &+\sum_n \sqrt{\frac{1}{\varepsilon_0 V}} \big[ \tilde{\eps}^{\lambda}_{k}(z) \cdot \hat{\bmu }_n + \sum_{a,b\in \{x,y,z\}} \hat{Q}_{ab,n} \nabla_{a,n} \tilde{\eps}^{\lambda}_{k,b}(z) \big] \hat{p}_{k} \\
    &-  \sum_n \sqrt{\frac{k^2}{\varepsilon_0 V}} \lambda \tilde{\eps}^{\lambda}_{k}(z) \cdot \hat{\textbf{m}}_n \hat{q}_{k}.
\end{split}
\end{equation}

The self-magnetization term mediated via $\hat{\chi}^m_{n,ij}$ represents our first obstacle since it combines operators to cubic order, thus going beyond the otherwise quadratic form. In combination with the self-polarization term, the self-magnetization ensures gauge invariance and the stability of the combined system which renders it essential for any future developments of \textit{ab initio} cavity QED. As we strive for a simple analytical solution, we will assume a parametric dependence $\hat{\chi}^m_{n,ij} \approx \chi^m_{n,ij}$ such that $\bar{\omega}_k^2 = \omega_k^2 \big[ 1 + 2\sum_n\sum_{i,j=1}^3 \chi^m_{n,ij} \tilde{\eps}^{\lambda}_{k,i}(z) \tilde{\eps}^{\lambda}_{k,j}(z) / (c^2 \varepsilon_0 V) \big]$ characterized the effective photonic frequency. While this correction is inherently small ($\propto 1/c^2$), it should be noted that it scales linear in the number of molecules. As $\chi^m_{n,ij}$ is rarely even mentioned and to the best of our knowledge never considered, it remains an open problem to specify its dynamic value. It is however possible to set the value for $\bar{\omega}_k$ in relation to the systems characteristic solution. Using the sum-rule $ \langle m \vert [\hat{p}_{k}, [\hat{H}, \hat{p}_{k}]] \vert m \rangle = 2 \sum_{n} (E_n-E_m) \vert \langle m\vert \hat{p}_{k} \vert n \rangle \vert^2 $ with eigenvalues $E_m$ and eigenstates $\vert m \rangle$ provides $\bar{\omega}_k^2 = 2\sum_{n} (E_n-E_m)/\hbar^2 \vert \langle m\vert \hat{p}_{k} \vert n \rangle \vert^2 \quad \forall~m$. We will formally retain the self-magnetization term but ultimately ignore its influence on the visualization. It should be noted that the following model features a magnetic instability for sizeable particle number and $\xi$ if $\bar{\omega}_k$ is not adjusted accordingly. We will limit our discussion to the stable domain, i.e., where phase transitions are absent. Importantly, any \textit{ab initio} calculation that includes the flexibility to change the electric/nuclear structure should include \textit{all} components \cite{schafer2019relevance,rokaj2017}.

\subsection{Transition dipole moments}

\Cadd{We will illustrate the identification and relation of the transition moment for the two-level approximation in the following. The corresponding Hopfield model, featuring an identification with harmonic oscillators instead, is derived analogously. We disregard permanent dipole moments for brevity}
\begin{align}
    \hat{\bmu}_n \rightarrow (\bmu^n_{10} \hat{\sigma}_+ +  \bmu ^{n}_{01} \hat{\sigma}_-),\\ 
    \hat{\bmag}_n \rightarrow (\bmag_{10}^n \hat{\sigma}_+ +  \bmag^{n}_{01} \hat{\sigma}_-),
\end{align}
where the matrix elements of the transition dipole moment (TDM) operators are calculated according to
\begin{equation}
    \bmu_{01} = \langle 0 \rvert \hat \bmu \rvert 1 \rangle,\ \ 
    \bmag_{01} = \langle 0 \rvert \hat \bmag \rvert 1 \rangle,
\end{equation}
and the lowering and raising operators of the TLS are given by the standard expressions
\begin{equation}
    \sigma_{+} = \rvert 1 \rangle \langle 0 \rvert,\ \ 
    \sigma_{-} = \sigma_{+}^{\dag} = \rvert 0 \rangle \langle 1 \rvert.
\end{equation}
Without the loss of generality, the matrix element of the electric TDM operator may be assumed real-valued, $\bmu_{01} = \bmu_{01}^*$.

\Cadd{
Let us establish the general relationship between transition dipole moments of a two-level quantum emitter.
For a bi-isotropic molecule with parallel electric and magnetic transition dipole moments this equation takes the simple form \cite{Condon1937}:
\begin{equation}
    \bmag^n_{01} = -i c \xi \bmu ^{n}_{01}, 
    \label{Eq_S12}
\end{equation}
with $\xi= \pm 1$ corresponding to LH ($+1$) and RH ($-1$) emitters, respectively.
Correspondingly, the magnetic dipole moment operator becomes $\hat \bmag = i c \xi (\bmu_{01}^* \hat \sigma_+ - \bmu_{01} \hat \sigma_-)$.
The above relationship between the TDMs of a chiral emitter is consistent with
the chirality definition of a classical monochromatic dipolar source \cite{corbaton2014helicity}.
Let us now look for a more general tensorial expression that would relate the transition dipole moments of an anisotropic molecular emitter.}

\begin{figure}[h!]
\includegraphics[width=.4\textwidth]{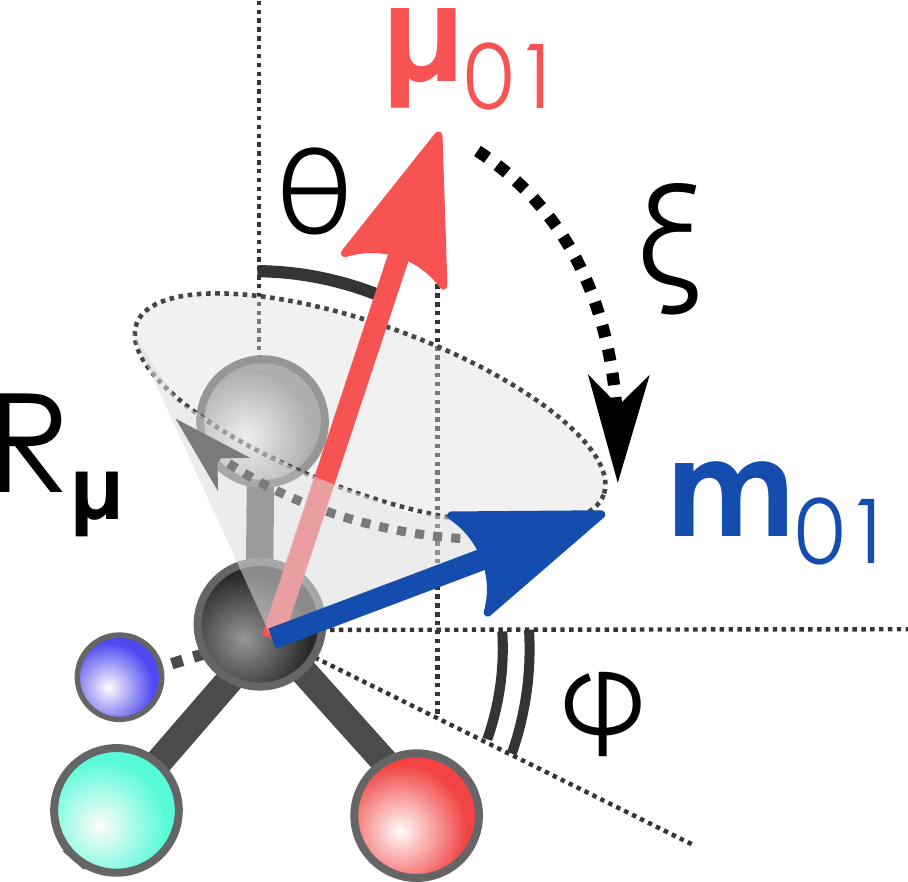}
\caption{Illustration of the relationship between the electric and magnetic transition dipole moments in a generic molecular emitter.
The orientation of the transition electric dipole moment $\bmu_{01}$ associated with the molecule can be described by two angles $\theta, \varphi$ in a local spherical coordinates system.
$\dyad{\xi}$ describes a unitary mapping $\bmu \to \bmag$. For a given orientation of the electric dipole moment $\bmu_{01}$, all allowed positions of $\bmag_{01}$ occupy a circle denoted by the shaded area. This additional mapping is accomplished by the rotation of the molecule around $\bmu_{01}$, which is described by $R_\mu$.}
\label{fig_S1}
\end{figure}

\Cadd{
Three Euler angles unambiguously describe the orientation of any rigid body (such as a molecule) in space. Similarly, a pair of non-collinear vectors $\bmu_{01}$ and $\bmag_{01}$ hard-wired to the molecule would work. But $\bmu_{01}$ and $\bmag_{01}$ are not independent themselves, and must satisfy a molecule-specific characteristic equation. 
First let us fix the orientation of the electric dipole moment of the molecule $\bmu_{01}$. This vector needs to be mapped into the magnetic transition dipole moment $\bmag_{01}$.
Suppose this mapping is performed by a dyad $\dyad{\xi}$, as illustrated in Fig. S2.
This dyad cannot be an arbitrary linear transformation: the action of $\dyad{\xi}$ on $\bmu_{01}$ must be invariant with respect to an arbitrary rotation of the molecule.
Thus, $\dyad{\xi}$ must be a product of an orthogonal linear transformation $\dyad{U}$ in $\mathbb{R}^3$ and scaling by a complex number $s$:
\begin{equation}
    \dyad{\xi} = s \cdot \dyad{U}.
    \label{Eq_S13}
\end{equation}
We can further specify its form by incorporating two rotations into the spatial orientation of $\bmu_{01}$.
Choosing a linearly-polarized transition electric dipole moment $\bmu_{01}$ is equivalent to fixing two Euler angles of the molecular orientation, leaving an arbitrary rotation around $\bmu_{01}$. For example, we can parameterize the electric TDM by the polar and azimuthal angles in the spherical coordinate system:
\begin{equation}
    \bmu_{01} = |\bmu|
    \left( \begin{array}{c}
    \sin \theta \cos \varphi \\ 
    \sin \theta \sin \varphi\\
    \cos \theta \end{array}
    \right).
    \label{Eq_S14}
\end{equation}
A given tensor $\dyad{\xi}$ will map a given electric dipole moment $\bmu_{01}$ into another fixed vector.
However, given a fixed $\bmu_{01}$ all allowed positions of $\bmag_{01}$ occupy an entire circle in $\mathbb{R}^3$. Thus, mapping $\bmu_{01} \to \bmag_{01}$ must be parameterized by an additional angle describing rotations of the molecule around $\bmu_{01}$ (see Fig.~\ref{fig_S1}). The sought for mapping thus can be written as:
\begin{equation}
    \bmag_{01} = -i c \dyad{R}_\mu (\delta) \dyad{\xi}\bmu_{01},
    \label{Eq_S15}
\end{equation}
where $\dyad{R}_\mu(\delta)$ is the rotation matrix that describes rotation of the molecule around $\bmu_{01}$ by an angle $\delta$.
The mapping parameterized by three angles ($\theta,\varphi,\delta$) in Eq. \ref{Eq_S15} encompasses all possible orientations of the molecule, which allows us to average any characteristic of the coupled system (such as the coupling constant) over molecular orientation.
}


Let us now establish the constraints imposed on $\dyad{\xi}$ by reciprocity.
The linear relationship between induced electric and magnetic dipole moments of a polarizable subwavelength object and incident monochromatic field can be written as:
\begin{equation}
    \left( \begin{array}{c}
    \bp  \\ 
    \bmag  \end{array}
    \right) = 
    \left( \begin{array}{cc}
    \eps_0\dyad{\alpha}_e &  \dyad{\alpha}_{em}/c \\ 
     \eps_0 c \dyad{\alpha}_{me} &  \dyad{\alpha}_m
    \end{array} \right)
    \left( \begin{array}{c}
    \bE \\ 
    \bH \end{array} \right),
    \label{Eq_S16}
\end{equation}
where $\dyad{\eps}$, $\dyad{\mu}$, $\dyad{\alpha}_{em}$, and $\dyad{\alpha}_{me}$ are all rank-2 $3 \times 3$ tensors with units of volume. 

We limit our treatment to the class of reciprocal media; polarizabilities of any reciprocal particle are subject to Onsager-Casimir relations \cite{caloz2018electromagnetic}:
\begin{equation}
\begin{split}
    \dyad{\alpha}_{e} = \dyad{\alpha}_{e}^T, \\
    \dyad{\alpha}_{m} = \dyad{\alpha}_{m}^T, \\
    \dyad{\alpha}_{em} = - \dyad{\alpha}_{me}^T.
\end{split}
    \label{Eq_S17}
\end{equation}
This criterion allows us to decompose the magneto-electric coupling tensors into reciprocal ('R') and non-reciprocal ('NR') components:
\begin{align}
     \dyad{\alpha}_{em} =  \dyad{\alpha}_{em}^{(NR)} + \dyad{\alpha}_{em}^{(R)} \equiv \dyad{\chi} + i \dyad{\kappa},\\
    \dyad{\alpha}_{me} = \dyad{\alpha}_{me}^{(NR)} + \dyad{\alpha}_{me}^{(R)} \equiv \dyad{\chi}^T - i \dyad{\kappa}^T.
\end{align} 
where the reciprocal part is presented by $\dyad{\kappa}$:
\begin{equation}
    \kappa = \frac{\dyad{\alpha}_{em} - \dyad{\alpha}_{me}^T}{2i},
    \label{Eq_S20}
\end{equation} 
and the non-reciprocal part is presented by
\begin{equation}
    \chi = \frac{\dyad{\alpha}_{em} + \dyad{\alpha}_{me}^T}{2}.
    \label{Eq_S21}
\end{equation} 
Obviously, if $\dyad{\chi} = 0$, then $\dyad{\alpha}_{em} = - \dyad{\alpha}_{me}^T$, and the reciprocity criterion is satisfied. It is self-explanatory that the reciprocal part of magneto-electric polarizability is responsible for effects that respect reciprocity.

Atomic polarizabilities of the elementary two-level system with arbitrary transition dipole moments $\bmu$ and $\bmag$ can be written as:
\begin{equation}
    \dyad{\alpha}_{e} = \frac{\bmu_{01} \otimes \bmu_{01}^*}{\hbar \eps_0} \frac{1}{\omega_0 - \omega - i \gamma /2},\quad  
    \dyad{\alpha}_{m} = \frac{\bmag_{01} \otimes \bmag_{01}^*}{\hbar \eps_0 c^2} \frac{1}{\omega_0 - \omega - i \gamma /2},
    \label{Eq_S22}
\end{equation}
\begin{equation}
    \dyad{\alpha}_{em} = \frac{ \bmu_{01} \otimes \bmag_{01}^*}{\hbar \eps_0 c} \frac{1}{\omega_0 - \omega - i \gamma /2},\quad  
    \dyad{\alpha}_{me} = \frac{ \bmag_{01} \otimes \bmu_{01}^*}{\hbar \eps_0 c} \frac{1}{\omega_0 - \omega - i \gamma /2}.
    \label{Eq_S23}
\end{equation}
Equations \ref{Eq_S22} suggest that for $\dyad{\alpha}_e$ and $\dyad{\alpha}_m$ to comply with Casimir-Onsager relations, Eq. \ref{Eq_S17}, $\bmu_{01}$ and $\bmag_{01}$ must be real-valued vectors (up to an arbitrary global phase $e^{i \phi}$), thus describing linearly-polarized transitions.
One can easily see that plugging $\bmag_{01} = -i c \xi \bmu_{01}$ (Eq. \ref{Eq_S12}) into the above expressions yields $\dyad{\alpha}_{em} = - \dyad{\alpha}_{me}^T$ and $\dyad{\chi} = 0$.

Now let us utilize Eq. \ref{Eq_S15} and for brevity work with the numerator of the full expression in Eq. \ref{Eq_S23}:
\begin{equation}
\begin{split}
    \dyad{\alpha}_{em} &\propto \bmu_{01} \otimes \bmag_{01}^* \equiv
    \bmu_{01} \bmag_{01}^{\dag} = 
    \bmu_{01} \left( -ic \dyad{R}_{\bmu} \dyad{\xi} \bmu_{01} \right)^{\dag} \\
    &= \bmu_{01} \left( -ic \bmu_{01}^T (\dyad{R}_{\bmu} \dyad{\xi})^T  \right)^{*} =
    ic \bmu_{01} \bmu_{01}^{\dag} (\dyad{R}_{\bmu} \dyad{\xi}) ^{\dag}.
\end{split}
\label{Eq_S24}
\end{equation}
Similarly, for $\alpha_{me}$ we obtain:
\begin{equation}
    \dyad{\alpha}_{me} \propto \bmag_{01} \otimes \bmu_{01}^* \equiv
    \bmag_{01} \bmu_{01}^{\dag} = 
     -ic  \left( \dyad{R}_{\bmu} \dyad{\xi} \right) \bmu_{01} \bmu_{01}^{\dag}  .
     \label{Eq_S125}
\end{equation}
Transposing the latter, assuming without the loss of generality the real-valued $\bmu_{01}$ and inserting into the Casimir-Onsager relation, we get:
\begin{equation}
    \dyad{\alpha}_{em} + \dyad{\alpha}_{me}^T \propto 
    ic \bmu_{01} \bmu_{01}^{\dag} [ ( \dyad{R}_{\bmu} \dyad{\xi})^{\dag} -( \dyad{R}_{\bmu} \dyad{\xi})^T ] = 0.
    \label{Eq_S26}
\end{equation}
Since $\dyad{R}_{\bmu}$ is a real-valued orthogonal matrix, the latter implies that for a reciprocal bi-anisotropic two-level emitter $\dyad{\xi}$ must be real-valued:
\begin{equation}
    \boxed{\Im[ \dyad{\xi}] = 0.}
    \label{Eq_S27}
\end{equation}
In other words, the scaling factor $s$ of the transformation is real-valued, $\Im[s] = 0$.

\Cadd{For now, let us assume that the transition dipole moments are related by the simple expression with a scalar $\xi$, Eq. \ref{Eq_S12}.}
Using this compact relation between electric and magnetic moments allows us to combine electric dipole, electric quadrupole and magnetic dipole into a single compact expression (assuming $\bmu ^n_{10}=\bmu ^{n,*}_{10}$ and $Q_{ab,n}^{10}=Q_{ab,n}^{01},~ \textbf{Q}_n = Q_{ab,n}^{10} \nabla_a \textbf{e}_b$):
\begin{equation}
\begin{split}
    \hat{H} &= \sum_n \hat{H}_{M,n} + \frac{1}{2\varepsilon_0 V}\Big[\sum_n \tilde{\eps}^{\lambda}_{k}(z) \cdot \bmu ^n_{01} (\sigma_n^+ +\sigma_n^-)\Big]^2
    + \hbar \bar\omega_k \left(\hat{a}^\dagger \hat{a} + \frac{1}{2}\right) \\
    &-i\sum_n \bar{g}_n \big[ (\sigma_n^+ +\sigma_n^-) (\hat{a}-\hat{a}^\dagger) - \bar{\xi}_n \lambda  (\sigma_n^+ -\sigma_n^-)(\hat{a}+\hat{a}^\dagger) \big]
\end{split}
\end{equation}
where
\begin{equation}
    \bar{g}_n = \sqrt{\frac{\hbar \bar{\omega}_k}{2\varepsilon_0 V}} (\bmu ^n_{10} + \textbf{Q}_n) \cdot \tilde{\eps}^{\lambda}_{k}(z)
\end{equation}
and
\begin{equation}
    \bar{\xi}_n = \frac{\omega_k}{\bar{\omega}_k} \frac{\bmu ^n_{10}\cdot \tilde{\eps}^{\lambda}_{k}(z)}{(\bmu ^n_{10} + \textbf{Q}_n) \cdot \tilde{\eps}^{\lambda}_{k}(z)} \xi.
\end{equation}

\subsection{Tavis-Cummings models -- a hint at the effective coupling strength}

Let us here introduce the common matter-representation of two-level models, discard the self-polarization term and all counter-rotating terms ($\propto \hat{a}\hat{\sigma}^- \approx 0$ and similar). We obtain the strongly simplified chiral Tavis-Cummings Hamiltonian:
\begin{align}
\label{eq:CTC}
    \hat{H}_{CTC} = \sum_n \hbar\omega_m \hat{\sigma}^+_n \hat{\sigma}^-_n + \hbar \bar{\omega}_k (\hat{a}^\dagger \hat{a} + \frac{1}{2}) 
- i\hbar \sum_n \bar{g}_n (1+\bar{\xi}_n \lambda) \big[ \hat{\sigma}^+_n \hat{a} - \hat{\sigma}^-_n \hat{a}^\dagger\big]
\end{align}
that shows explicitly that the effective interaction strength is proportional to $\bar{g}_n (1+\bar{\xi}_n \lambda)$. A chiral emitter that features the same handedness as the cavity will couple stronger to the mode. In the extreme case that $\bar{\xi}_n = \pm 1$, the mismatched enantiomer will entirely decouple from the mode. The above chiral Tavis-Cummings model could be  solved analytically in the same way as any Tavis-Cummings model, i.e, by limiting ourselves to the single-excitation subspace and introducing collective spin-operators. Here, we will focus on the Hopfield model that includes also the counter-rotating and self-polarization terms that can lead to sizeable renormalizations for large $N$.

\subsection{Chiral Hopfield model}
\Cadd{In contrast to the Dicke and Tavis-Cummings models with a two-level approximation, the Hopfield approach is based on representing the material in terms of harmonic oscillators which allows for an analytic solution also in the ultra-strong coupling domain. The first polariton manifold is identical in the single-excitation + strong-coupling regime. Our Chiral Hopfield Hamiltonian takes the form of $N+1$ coupled harmonic oscillators}
\begin{equation}
\begin{split}
    \hat{H} &= \sum_n \hbar \omega_n (\hat{b}^\dagger_n \hat{b}_n + \frac{1}{2}) + \hbar \bar{\omega}_k (\hat{a}^\dagger \hat{a} + \frac{1}{2}) \\
    &- i\hbar \sum_n \bar{g}_n \big[ (\hat{b}^\dagger_n + \hat{b}_n) (\hat{a} - \hat{a}^\dagger)
    + \bar{\xi}_n \lambda (\hat{b}^\dagger_n - \hat{b}_n) (\hat{a} + \hat{a}^\dagger) \big]\\
    &+\frac{1}{2\varepsilon_0 V}\Big[\sum_n \tilde{\eps}^{\lambda}_{k}(z) \cdot \bmu ^n_{01} (\hat{b}_n^\dagger +\hat{b}_n)\Big]^2.
\end{split}
\end{equation}
\Cadd{The transition-dipole moments are related to the optical oscillator strength of the harmonic model.}
Assuming identical molecules, it is convenient to introduce the Fourier-representation for the molecular ensemble
\begin{align}
    \hat{B}_{\bk}^\dagger = \frac{1}{\sqrt{N}} \sum_{n} e^{i \bk \cdot \br_n} \hat{b}_n^\dagger,~ \hat{b}_n^\dagger = \frac{1}{\sqrt{N}} \sum_{\bk} e^{-i \bk \cdot \br_n} \hat{B}_{\bk}^\dagger
\end{align}
such that the collective operators are represented by bright $\sum_n \hat{b}^\dagger = \sqrt{N} \hat{B}_{k=0}^\dagger$ and dark states
\begin{equation}
\begin{split}
    \hat{H} &\approx  \hbar\omega_m (\hat{B}_{k=0}^\dagger\hat{B}_{k=0} + \frac{1}{2}) + \hbar\omega_m \sum_{k\neq 0}\hat{B}_{k}^\dagger\hat{B}_{k} + (N-1)\frac{\hbar \omega_m}{2} + \hbar\bar{\omega}_k (\ha ^\dagger\ha +\frac{1}{2})\\
    &- i \hbar \sqrt{N}\bar g \big[ (\hat{B}_{k=0}^\dagger + \hat{B}_{k=0}) (\ha  - \ha ^\dagger) + \bar \xi \lambda  (\hat{B}_{k=0}^\dagger - \hat{B}_{k=0})(\ha  + \ha ^\dagger) \big] \\
    &+ \frac{N}{2\varepsilon_0 V}\Big[\tilde{\eps}^{\lambda}_{k}(z) \cdot \bmu _{01} (\hat{B}_{k=0}^\dagger +\hat{B}_{k=0})\Big]^2~.
\end{split}
\label{eq:hopfield}
\end{equation}
Disregarding the dark states $\hbar\omega_m \sum_{k\neq 0}\hat{B}_{k}^\dagger\hat{B}_{k}$ and their vacuum-fluctuations $(N-1)\frac{\hbar \omega_m}{2}$, the self-polarization term can be absorbed into an adjusted matter frequency of the bright states
$ \tilde{\omega}_m^2 = \omega_m^2 + N \frac{2\omega_m}{\hbar \varepsilon_0 V} (\tilde{\eps}^{\lambda}_{k}(z) \cdot \bmu_{01})^2 $ which results in
\begin{equation}
\begin{split}
    \hat{H} &=  \hbar\tilde\omega_m (\hat{B}_{k=0}^\dagger\hat{B}_{k=0} + \frac{1}{2}) + \hbar\bar{\omega}_k (\ha ^\dagger\ha +\frac{1}{2})\\
    &- i \hbar \sqrt{N}\tilde g \big[ (\hat{B}_{k=0}^\dagger + \hat{B}_{k=0}) (\ha  - \ha ^\dagger)
    + \tilde \xi \lambda  (\hat{B}_{k=0}^\dagger - \hat{B}_{k=0})(\ha  + \ha ^\dagger) \big]~.
\end{split}
\label{eq:hopfield} 
\end{equation}
where $\tilde{g} = \sqrt{\frac{\hbar \bar{\omega}_k \omega_m}{2\varepsilon_0 V \tilde{\omega}_m}} (\bmu_{01} + \textbf{Q}_n) \cdot \tilde{\eps}^{\lambda}_{k}(z)$ and $\tilde{\xi} = \frac{\tilde{\omega}_m \omega_k}{\omega_m \bar{\omega}_k} \frac{\bmu_{01}\cdot \tilde{\eps}^{\lambda}_{k}(z)}{(\bmu_{01} + \textbf{Q}) \cdot \tilde{\eps}^{\lambda}_{k}(z)} \xi$ are the renormalized effective coupling strength and chirality factor.

We can diagonalize Eq.~\eqref{eq:hopfield} by following the standard Hopfield \cite{hopfield1958,todorov2012} procedure, i.e., defining the polaritonic operator $\hat{\Pi} = x \ha  + y \ha ^\dagger + z \hat{B}_{k=0}  + u \hat{B}_{k=0}^\dagger$  that fulfills the eigenvalue equation $[\hat{H},\hat{\Pi}] = \hbar \Omega \hat{\Pi}$ with the normalization condition $\vert x \vert^2 - \vert y \vert^2 + \vert z \vert^2 - \vert u \vert^2 = 1$.  
We obtain the polaritonic frequencies from 
\begin{align}
    \begin{vmatrix}
    - \bar{\omega}_k - \Omega & 0 & (1+\tilde\xi\lambda )i\sqrt{N}\tilde{g}& -(1-\tilde\xi \lambda )i\sqrt{N}\tilde{g}\\
    0 & \bar{\omega}_k -\Omega & -(1-\tilde\xi \lambda )i\sqrt{N}\tilde{g} & (1+\tilde\xi \lambda )i\sqrt{N}\tilde{g}\\
    -(1+\tilde\xi \lambda )i\sqrt{N}\tilde{g} & -(1-\tilde\xi \lambda )i\sqrt{N}\tilde{g} & -\tilde{\omega}_m - \Omega & 0\\
    -(1-\tilde\xi \lambda )i\sqrt{N}\tilde{g} & -(1+\tilde\xi \lambda )i\sqrt{N}\tilde{g} & 0 & \tilde{\omega}_m - \Omega
    \end{vmatrix}
    = 0
\end{align}
as real and positive solutions
\begin{widetext}
\begin{align}
    \Omega_{\pm} = \frac{1}{\sqrt{2}}\sqrt{ \bar{\omega}_k^2 + \tilde{\omega}_m^2 + 8 \tilde{\xi}\lambda  N \tilde{g}^2 \pm \sqrt{[\omega_k^2 - \tilde{\omega}_m^2]^2 + 16 N \tilde{g}^2 (\bar{\omega}_k + \tilde\omega_m \tilde\xi \lambda  ) (\bar{\omega}_k \tilde\xi \lambda  + \tilde\omega_m)  } }~.
\end{align}
\end{widetext}
The corresponding eigenvectors encode in $\vert x \vert^2 - \vert y \vert^2$ the matter contribution and in $\vert z \vert^2 - \vert u \vert^2$ the photonic contribution to the polaritonic states.

\subsection{Generic alignment and its influence on chiral recognition}

\Cadd{Let us briefly examine the more general case where the electric and magnetic transition dipole moments are arbitrarily oriented. 
In this case, they are related by Eq. \ref{Eq_S22} with  $\dyad{\xi} = s \dyad{U}$ where $s$ is real-valued and $\dyad{U}$ is an orthogonal transformation.}


\Cadd{
The orientation average is described by the energy-conserving squared coupling element:
\begin{align}
    \langle |g|^2 \rangle \propto 
    \frac{1}{4\pi} \int_0^{2\pi} d\phi  \int_0^\pi d\theta \sin(\theta) \vert \sqrt{N} \tilde{\varepsilon} \cdot (1+\lambda \dyad{\xi}) \bmu_{01} \vert^2
\end{align}
proportional to the combined electric plus magnetic moments.
Expanding the dot product
\begin{equation}
\begin{split}
    \vert \tilde\varepsilon \cdot  (1+\lambda \dyad{\xi}) \bmu_{01}  \vert^2
    =& \cos^2 \theta \vert (1+\lambda \dyad{\xi}) \bmu_{01} \vert^2\\
    =& \cos^2 \theta \left( \vert \bmu_{01} \vert^2 + \langle \dyad{\xi} \bmu_{01} \vert \dyad{\xi} \bmu_{01}\rangle + 2 \lambda \langle \bmu_{01} \vert \dyad{\xi} \bmu_{01} \rangle \right)
\end{split}
    \label{eq:Sangle}
\end{equation}
and utilizing $\dyad{\xi} = s \dyad{U}$, we obtain:
\begin{equation}
\begin{split}
\label{eq:angularaverage}
    & \langle \vert g\vert^2 \rangle 
    = \lvert \sqrt{\frac{\hbar \bar{\omega}_k \omega_m }{ 2\varepsilon_0 V \tilde{\omega}_m }} \frac{\tilde{\omega}_m \omega_k}{\omega_m \bar{\omega}_k} \sqrt{N}(1+\lambda\dyad{\xi}) \cdot \bmu_{01} \rvert^2  
    \frac{1}{2} \int_0^\pi d\theta \sin(\theta) \cos^2(\theta) \\
    &= \frac{N}{3} \frac{\hbar \tilde{\omega}_m \omega_k^2 }{2 \varepsilon_0 V \bar{\omega}_k \omega_m} \vert (1+\lambda\dyad{\xi}) \cdot \bmu_{01} \vert^2 \\
    &= \frac{N}{3} \frac{\hbar \tilde{\omega}_m \omega_k^2 }{2 \varepsilon_0 V \bar{\omega}_k \omega_m} 
    \left[( 1+s^2)\vert \bmu_{01} \vert^2 + 2 \lambda s \Re \langle \bmu_{01} \vert \dyad{U} \bmu_{01} \rangle \right].
\end{split}
\end{equation}
}

\Cadd{ In addition to chiral features that arise from the parallel components of the transition dipole moments, the emitters now also feature Omega-type magneto-electric coupling originating from the orthogonal components of the dipole moments. However, as we can easily see from the angular average, only the chiral components are discriminated by the cavity.
Eq.~\eqref{eq:angularaverage} clarifies that the chiral cavity will only distinguish the chiral components of the emitters. Take for example pure chirality with $\lambda=+1$ and (anti-)alignment $\boldsymbol\mu || \textbf{m} $ with $\dyad{\xi} = \pm 1$, then 
$[(1+s^2)\vert \bmu_{01} \vert^2 + 2 \lambda s \Re \langle\bmu_{01} \vert \dyad{U}_\xi \bmu_{01}\rangle] = (1+1+ 2(\pm 1))\vert\bmu_{01}\vert^2$, which is either 0 or $4 \vert\bmu_{01}\vert^2 $. 
However, for $\Omega$-coupling $\langle\bmu_{01} \vert \dyad{U} \bmu_{01} \rangle = 0$ and we obtain always $(1+s^2) \vert\bmu_{01}\vert^2 $. The transition-dipole moments still contribute constructively to the chiral coupling but there is no handedness selectivity left.
}

\subsection{Influence of the self-magnetization}
\Cadd{
Let us illustrate briefly how the self-magnetization can influence our conclusions. First of all, it should be noticed that the dressing of the photonic frequency via $\chi^m$ is a factor $1/c^2$ smaller than the self-polarization effect on the matter frequency. However, even if we chose enormously large values, the effect is small as demonstrated in Fig.~\ref{fig:chimag}.\\
\begin{figure}[h!]
\includegraphics[width=0.5\textwidth]{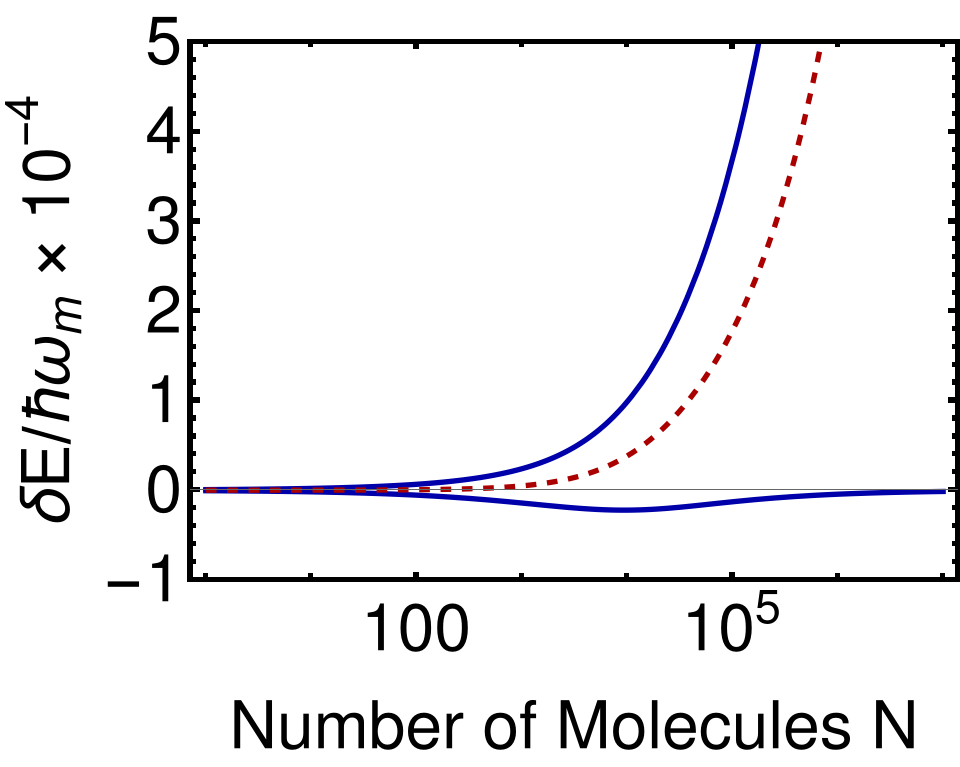}%
\includegraphics[width=0.5\textwidth]{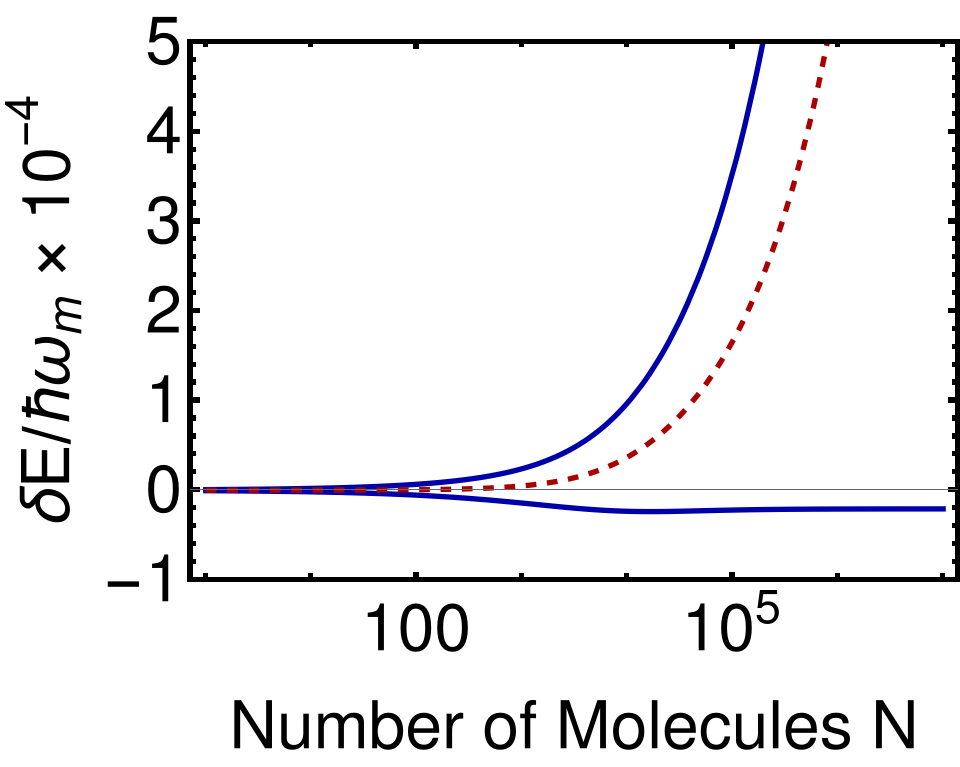}  
\caption{Left, including a quite large self-magnetization of $\chi^m = \mu^2$. Right, using a vastly enhanced self-magnetization of $\chi^m = c^2 \mu^2$, so even if we compensate the $1/c^2$ factor, the effect of the self-magnetization is small. Interestingly, the large self-magnetization can even enhance the recognition capabilities in the lower polariton.}
    \label{fig:chimag}
\end{figure}
}

\subsection{Extension to modes with non-zero in-plane momentum $k_x \neq 0$}

\Cadd{Our previous derivations used the simplifying assumption of a cavity mode represented by a standing wave with $\bk = \pm k \textbf{e}_z$, which leads to compact and highly intuitive equations. A more generic description might allow for non-zero in-plane momentum $k_x \neq 0$, resulting in mixing of propagating waves and dark states.
We will provide in the following a brief discussion what such an extension would look like and what changes are to be expected. We would like to emphasize that such an extended model would go more naturally with a many-mode description and present a straightforward generalization of our work.
}


\Cadd{
A planar optical cavity, such as the one described in ref. \cite{Voronin2022}, supports a continuous spectrum of resonant states that can be labeled by their in-plane momenta $\bf k_{\parallel}$. 
More sophisticated cavities, such as micro-domes, support more complex modes with in-plane contribution and non-Gaussian spot-distribution but we retain here with the simplified Fabry-P\'erot set up. 
A minimal representation for the field of such cavity modes is given in Section \textit{Fields of a standing chiral wave}.
Cavity modes with $\bf k_{\parallel} \ne 0$ do maintain their single-handedness quality in a substantial range of in-plane wave vectors (incident angles) according to the findings of ref. \cite{Voronin2022}, and thus are expected to feature similar energy spectra when coupled with chiral molecular emitters.
}

\Cadd{As before, we can expand the fields in its eigenmodes
\begin{align*}
    \hat{\bD}^\lambda_\perp(\br) &= i \sum_{\bk} \sqrt{\frac{\hbar c k \eps_0}{2 V}} \big( \epsilon_{\bk\lambda}(\br) \ha _{\bk\lambda} - \epsilon_{\bk\lambda}^*(\br) \ha ^\dagger_{\bk\lambda} \big),\\
    \hat{\bB}^\lambda(\br) &= i \frac{1}{c} \sum_{\bk} \sqrt{\frac{\hbar c k }{2 \eps_0 V}} \big( \beta_{\bk\lambda}(\br) \ha _{\bk\lambda} - \beta_{\bk\lambda}^*(\br) \ha ^\dagger_{\bk\lambda} \big)
\end{align*}
where $\epsilon_{\bk\lambda}(\br) = \frac{1}{\sqrt{2}} (k_z/k, i\lambda, -k_x/k)^T e^{ik_z z + ik_x x} $ and $\beta_{\bk\lambda}(\br) = -i\lambda\epsilon_{\bk\lambda}(\br)$ such that
\begin{align*}
    \hat{\bB}^\lambda(\br) &= \frac{1}{c} \sum_{\bk} \lambda \sqrt{\frac{\hbar c k }{2 \eps_0 V}} \big( \epsilon_{\bk\lambda}(\br) \ha _{\bk\lambda} + \epsilon_{\bk\lambda}^*(\br) \ha ^\dagger_{\bk\lambda} \big).
\end{align*}
We again restrict enforce the chiral standing wave in z-direction by $\hat{\bD}^\lambda_\perp(\br) = \tfrac{1}{\sqrt{2}}( \hat{\bD}^\lambda_{\perp,k_z>0}(\br) + \hat{\bD}^\lambda_{\perp,k_z<0}(\br)) $ which results with $\ha _{k_x, k_z>0,\lambda} = \ha _{k_x, k_z<0,\lambda}$ in 
\begin{align*}
    \hat{\bD}^\lambda_\perp(\br) &= i \sum_{k_z>0,k_x} \sqrt{\frac{\hbar c k \eps_0}{2 V}} \big( \tilde\epsilon_{\bk\lambda}(\br) \ha _{\bk\lambda} - \tilde\epsilon_{\bk\lambda}^*(\br) \ha ^\dagger_{\bk\lambda} \big),\\
    \hat{\bB}^\lambda(\br) &=  \frac{1}{c} \sum_{k_z>0,k_x} \lambda \sqrt{\frac{\hbar c k }{2 \eps_0 V}} \big( \tilde\epsilon_{\bk\lambda}(\br) \ha _{\bk\lambda} + \tilde\epsilon_{\bk\lambda}^*(\br) \ha ^\dagger_{\bk\lambda} \big)
\end{align*}
where $\tilde\epsilon_{\bk\lambda}(\br) = (k_z/k \cos(k_z z), -\lambda \sin(k_z z), -ik_x/k \sin(k_z z))^T e^{ik_x x}$.\\
The projection of the matter degrees of freedom follows as before and the overall structure remains unchanged. As an example, we will derive the new Tavis-Cummings analogue, the Hopfield model can be obtain in analogy to the previous steps but the many-mode coupling renders the process more verbose. As before, we perform the rotating-wave approximation for the chiral Tavis-Cummings model and disregard all self-correction terms as well as quadrupole contributions to obtain
\begin{align*}
    \hat{H}^{\lambda}_{CTC} &= \sum_n \hbar\omega_m \hat{\sigma}^+_n \hat{\sigma}^-_n + \sum_{k_z>0,k_x} \hbar \omega_\bk (\hat{a}_{\bk\lambda}^\dagger \hat{a}_{\bk\lambda} + \frac{1}{2}) \\
    &- \sum_{n, k_z>0,k_x} i\sqrt{\frac{\hbar c k}{2 \eps_0 V}} \big[ (\tilde\epsilon_{\bk\lambda}(\br) \cdot \boldsymbol\mu_{01}^n \hat{\sigma}^+_n \hat{a}_{\bk\lambda} - h.a. ) + \lambda (\tilde\epsilon_{\bk\lambda}(\br) \dyad{\xi}_n \boldsymbol\mu_{01}^n \hat{\sigma}^+_n \hat{a}_{\bk\lambda} - h.a.) \big].\\
    &= \sum_n \hbar\omega_m \hat{\sigma}^+_n \hat{\sigma}^-_n + \sum_{k_z>0,k_x} \hbar \omega_\bk (\hat{a}_{\bk\lambda}^\dagger \hat{a}_{\bk\lambda} + \frac{1}{2}) 
    - \sum_{n, k_z>0,k_x} i\sqrt{\frac{\hbar c k}{2 \eps_0 V}} \big[ \tilde\epsilon_{\bk\lambda}(\br) ( 1 + \lambda \dyad{\xi}_n ) \cdot \boldsymbol\mu_{01}^n \hat{\sigma}^+_n \hat{a}_{\bk\lambda} - h.a. \big].
\end{align*}
We introduce again a Fourier-representation $\hat{\sigma}^+_n = \frac{1}{\sqrt{N}}\sum_{\textbf{K}} e^{-i\textbf{K} \cdot \br_n} \hat{S}^\dagger_\textbf{K}$ which implies a regular molecular distance along the x-axis $ \br_n = \textbf{e}_x \frac{2\pi n}{N} $ and assume identical couplings and frequencies. As $\sum_n e^{ik_x x_n} e^{-iK_x x_n} = N \delta(k_x-K_x) $, the chiral Tavis-Cummings Hamiltonian simplifies to
\begin{align}
\begin{split}
    \hat{H}^{\lambda}_{CTC} &= \sum_K \hbar\omega_m \hat{S}^+_K \hat{S}^-_K + \sum_{k_z>0,k_x} \hbar \omega_\bk (\hat{a}_{\bk\lambda}^\dagger \hat{a}_{\bk\lambda} + \frac{1}{2}) \\
    &- \sum_{k_z>0,k_x} i\sqrt{N} \sqrt{\frac{\hbar c k}{2 \eps_0 V}} \big[ \tilde\epsilon_{\bk\lambda}(z,x=0) ( 1 + \lambda \dyad{\xi} ) \cdot \boldsymbol\mu_{01} \hat{S}^+_{k_x} \hat{a}_{\bk\lambda} - h.a. \big].
\end{split}
\end{align}
Comparison with Eq.~\eqref{eq:CTC} clarifies that the chiral effect, i.e., $(1+\lambda \dyad{\xi})$, remains unchanged. An important difference is that for non-zero in-plane momentum, the coupling is not only to the mode $K_x = 0$ but also to higher momenta, similar to the standard Tavis-Cummings model as shown for example in Ref.~\cite{tichauer2021multi}.
}

\newpage

\section{Chiral Hopfield model under the assumption of cancelling instantaneous intermolecular contributions}

We provide here a derivation of the chiral Hopfield model under the assumption that the instantaneous intermolecular interactions cancel. Using the explicit form of the chiral fields, our starting point reads then
\begin{equation}
\begin{split}
    \hat{H} &= \sum_n \big[ \hat{H}_{M,n} + \frac{1}{2\varepsilon_0 V}(\tilde{\eps}^{\lambda}_{k}(z) \cdot \hat{\bmu }_n)^2 \big]\\
    &+ \frac{1}{2}\hat{p}_{k}^2 + \frac{1}{2}\big[ \omega_{k}^2 + 2\sum_n \hat{\chi}^m_{n,ij} \frac{k^2}{\varepsilon_0 V} \tilde{\eps}^{\lambda}_{k,i}(z) \tilde{\eps}^{\lambda}_{k,j}(z) \big] \hat{q}_{k}^2 \\
    &+ \sum_n \sqrt{\frac{1}{\varepsilon_0 V}} \big[ \tilde{\eps}^{\lambda}_{k}(z) \cdot \hat{\bmu }_n + \sum_{a,b\in \{x,y,z\}} \hat{Q}_{ab,n} \nabla_{a,n} \tilde{\eps}^{\lambda}_{k,b}(z) \big] \hat{p}_{k} \\
    &-  \sum_n \sqrt{\frac{k^2}{\varepsilon_0 V}} \lambda \tilde{\eps}^{\lambda}_{k}(z) \cdot \hat{\textbf{m}}_n \hat{q}_{k}
\end{split}
\end{equation}
where the only difference is that the self-polarization is local only.

We can follow the same steps as before and absorb the self-polarization into adjusted local matter frequencies
$ \tilde{\omega}_n^2 = \omega_n^2 + \frac{2m\omega_n}{\hbar \varepsilon_0 V} (\tilde{\eps}^{\lambda}_{k}(z) \cdot \bmu ^n_{10})^2 $ -- notice the missing N.
We obtain
\begin{equation}
\begin{split}
    \hat{H} &= \sum_n \hbar \tilde{\omega}_n (\hat{b}^\dagger_n \hat{b}_n + \frac{1}{2}) + \hbar \bar{\omega}_k (\hat{a}^\dagger \hat{a} + \frac{1}{2}) \\
    &- i\hbar \sum_n \tilde{g}_n \big[ (\hat{b}^\dagger_n + \hat{b}_n) (\hat{a} - \hat{a}^\dagger) + \tilde{\xi}_n \lambda (\hat{b}^\dagger_n - \hat{b}_n) (\hat{a} + \hat{a}^\dagger) \big]
\end{split}
\end{equation}
where as before $\tilde{g}_n = \sqrt{\frac{\hbar \bar{\omega}_k \omega_n}{2\varepsilon_0 V \tilde{\omega}_n}} (\bmu ^n_{10} + \textbf{Q}_n) \cdot \tilde{\eps}^{\lambda}_{k}(z)$ and $\tilde{\xi}_n = \frac{\tilde{\omega}_n \omega_k}{\omega_n \bar{\omega}_k} \frac{\bmu ^n_{10}\cdot \tilde{\eps}^{\lambda}_{k}(z)}{(\bmu ^n_{10} + \textbf{Q}_n) \cdot \tilde{\eps}^{\lambda}_{k}(z)} \xi$ are the renormalized effective coupling strength and chirality factor.
Introducing the Fourier-representation leads to
\begin{align}
    \label{eq:hopfield2}
    \begin{split}
    \hat{H} &=  \hbar\tilde\omega_m (\hat{B}_{k=0}^\dagger\hat{B}_{k=0}+\frac{1}{2}) + \hbar\tilde\omega_m \sum_{k\neq 0}\hat{B}_{k}^\dagger\hat{B}_{k} + (N-1)\frac{\hbar \tilde{\omega}_m}{2} + 2\hbar\bar{\omega}_k (\ha ^\dagger\ha +\frac{1}{2})\\
    &- i \hbar \sqrt{N}\tilde g \big[ (\hat{B}_{k=0}^\dagger + \hat{B}_{k=0}) (\ha  - \ha ^\dagger) + \tilde \xi \lambda  (\hat{B}_{k=0}^\dagger - \hat{B}_{k=0})(\ha  + \ha ^\dagger) \big]~.
    \end{split}
\end{align}
Importantly, also the dark states are now dressed by the self-polarization and $\tilde{\omega}_m$ does not depend on the number of molecules N. The analytic solution has the same form but deviates in $\tilde{\omega}_m$ which results in instabilities for large N.

\begin{figure}[h]
    \centering
    \includegraphics[width=0.5\textwidth]{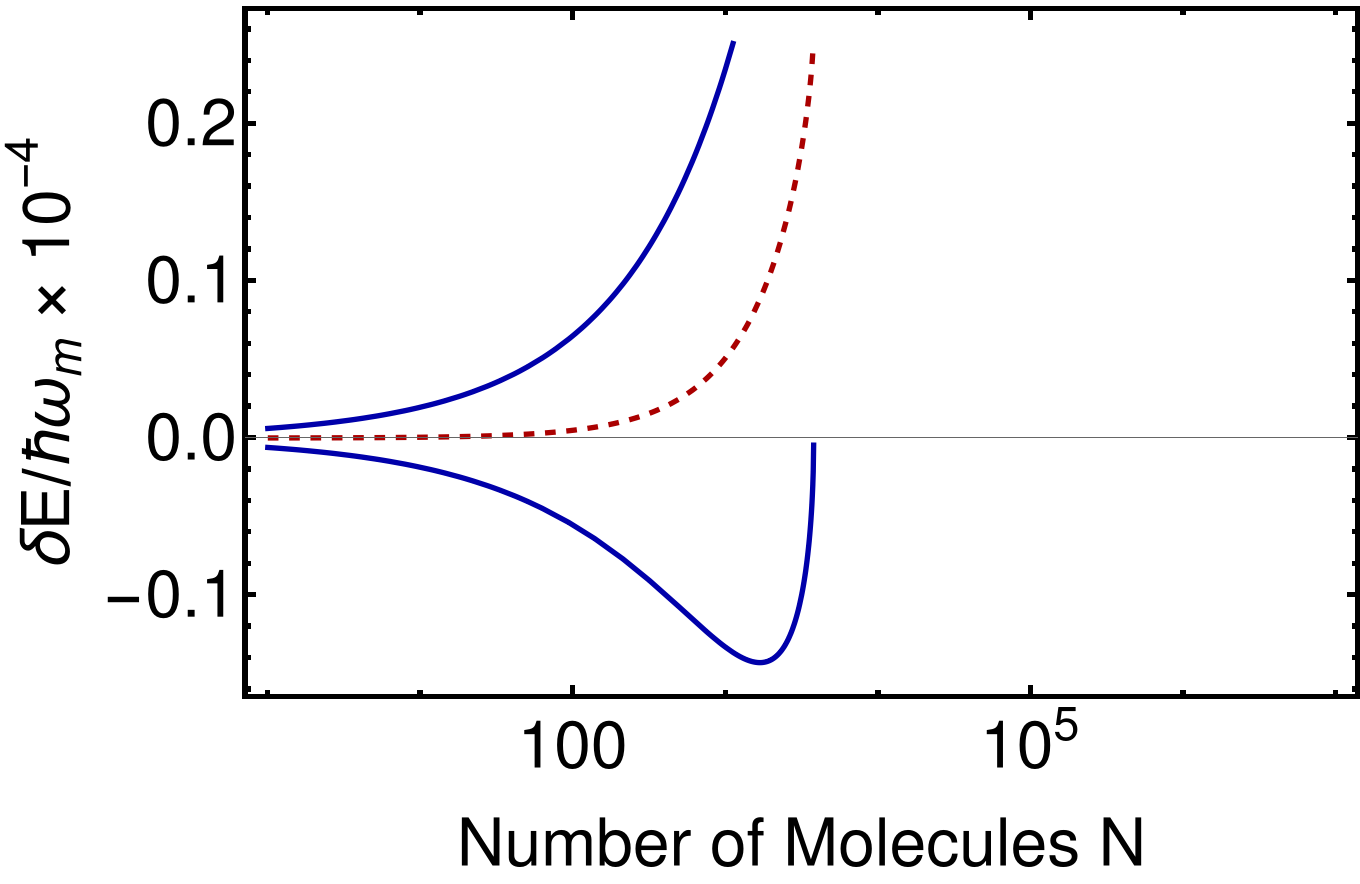}%
    \includegraphics[width=0.5\textwidth]{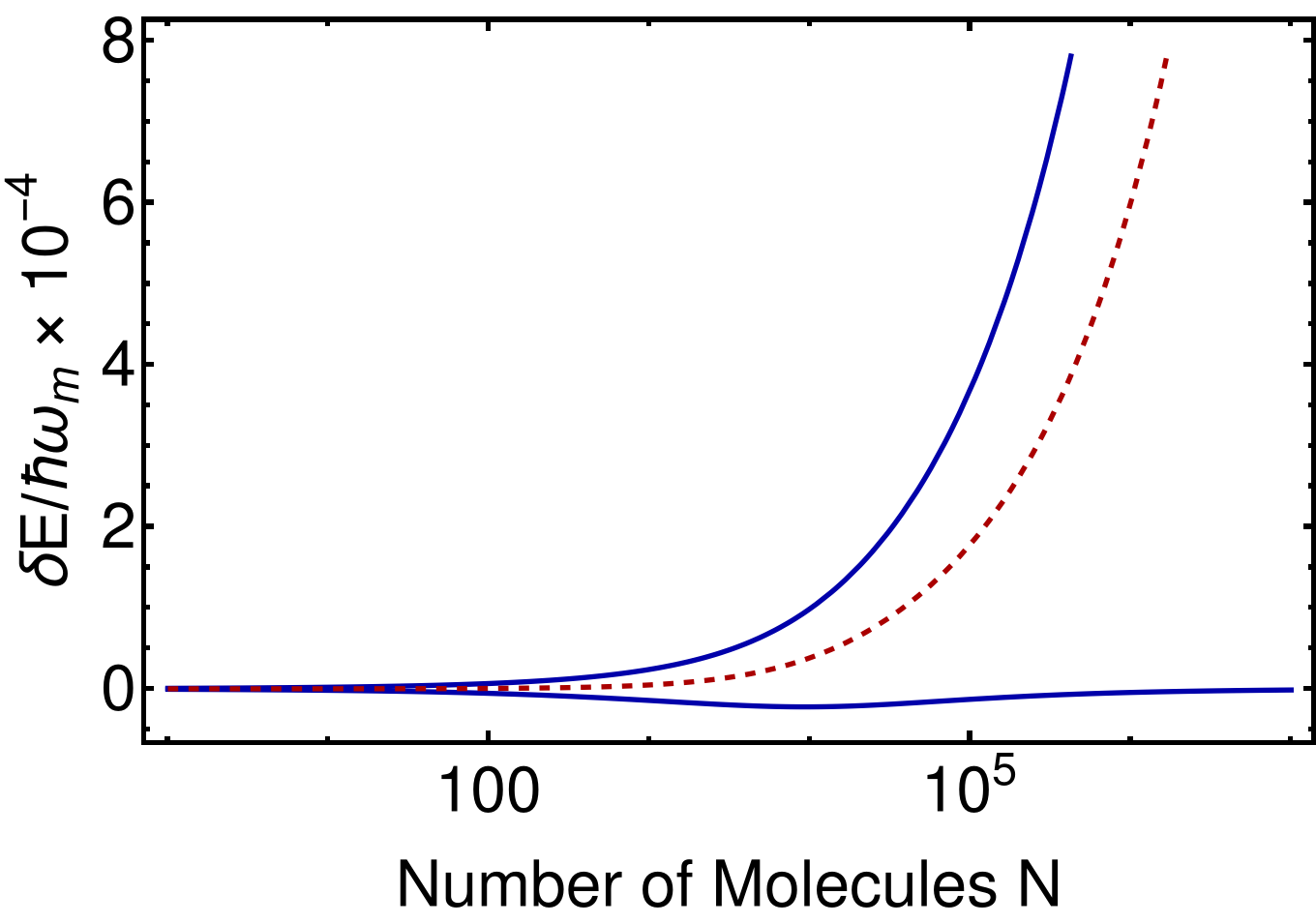}
    \caption{N-scaling differences between left and right-handed chiral dye molecules as introduced in the paper. Left, without intersystem self-polarization, right, using full self-polarization and disregarding intermolecular Coulomb interactions (same as in main text). All parameters consistent with Fig.~3.}
    \label{fig:diffeigs}
\end{figure}

Fig.~\ref{fig:diffeigs} contrasts the two version (left without intersystem self-polarization) of the chiral Hopfield model and the associated instability.

\clearpage
\newpage

\section{Consistency of the Chiral Hopfield model in the ultra-strong coupling domain with \textit{ab intio} calculations}

\begin{figure}[h]
    \centering
    \includegraphics[width=0.5\textwidth]{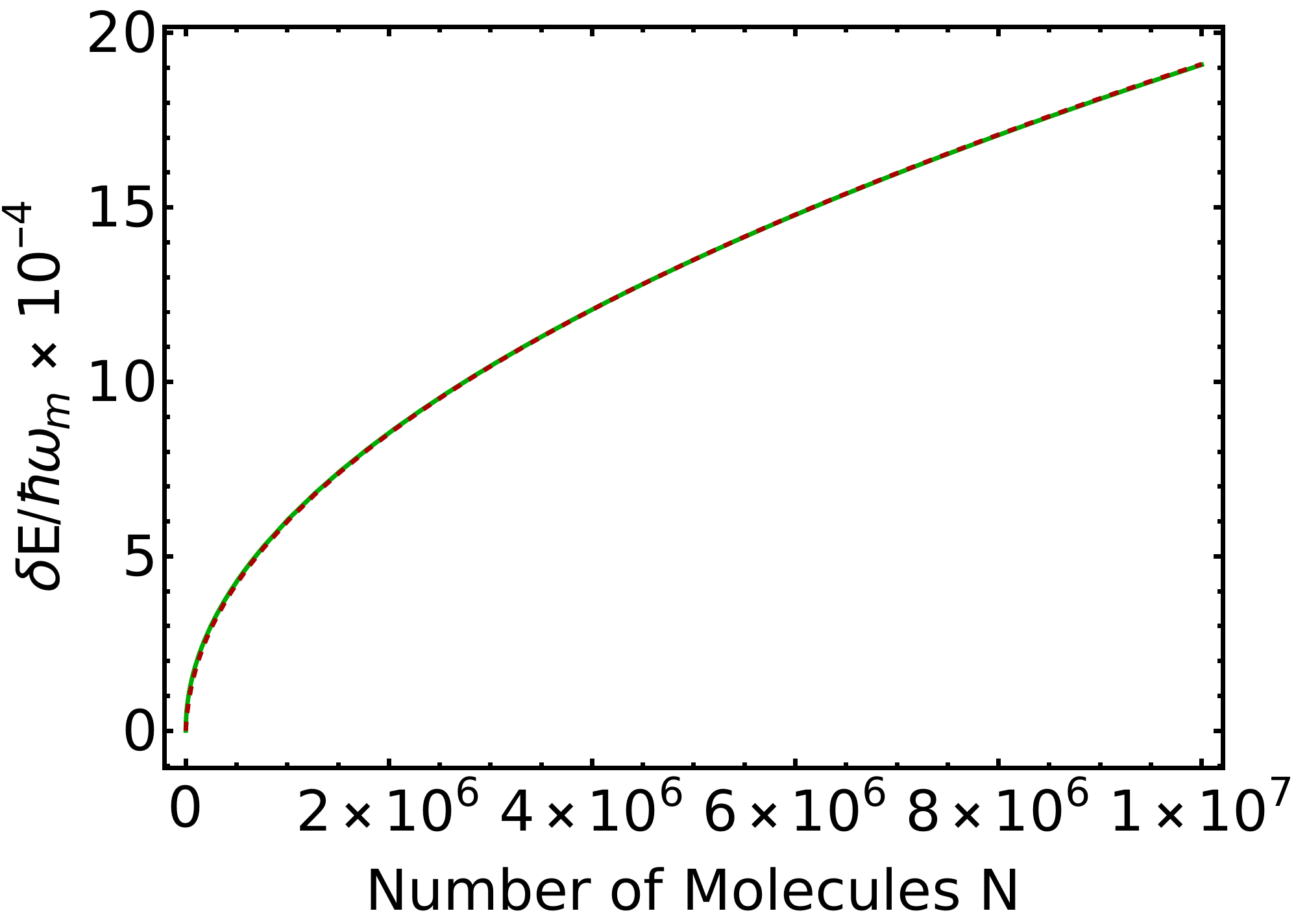}%
    \includegraphics[width=0.5\textwidth]{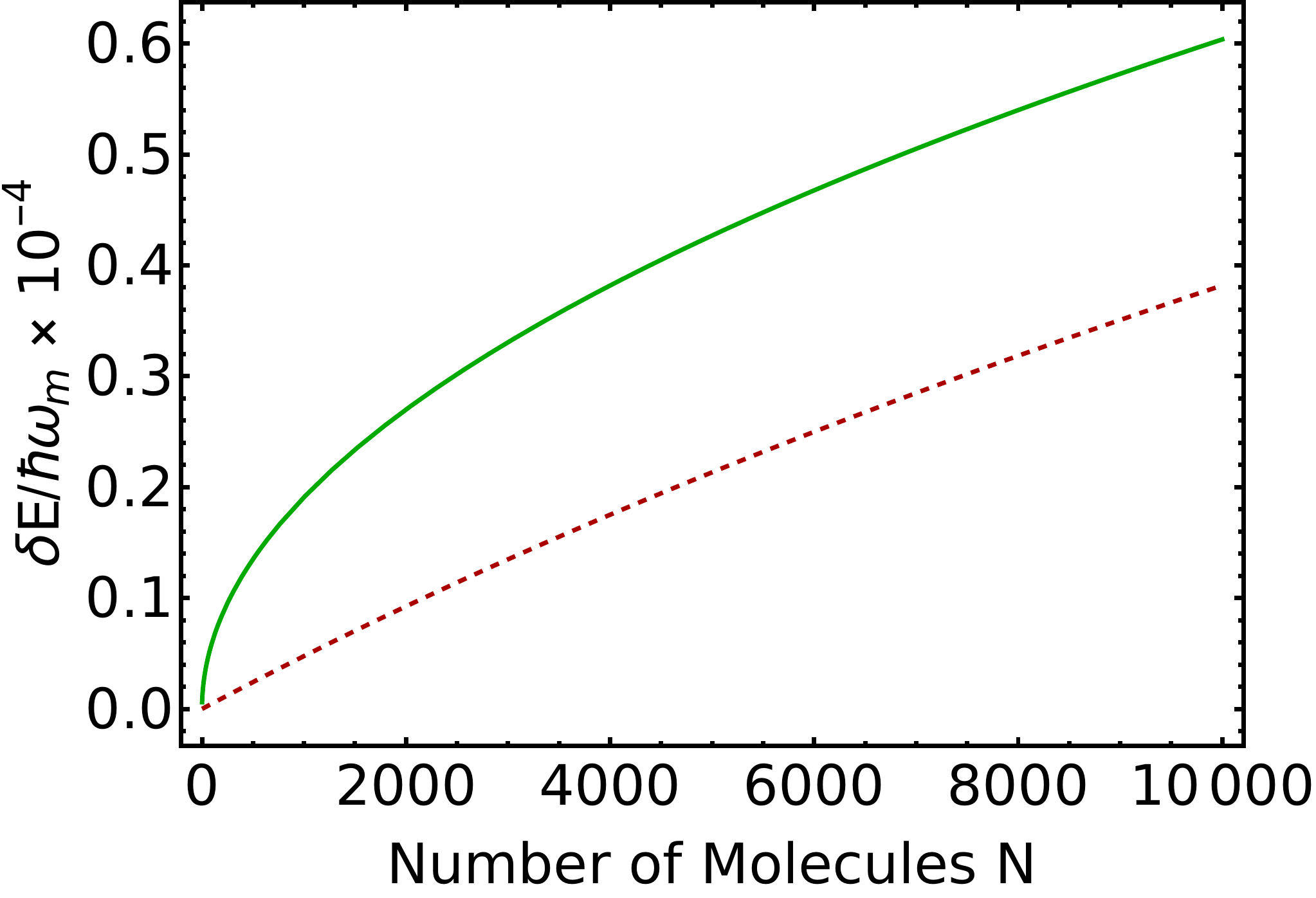}
    \caption{N-scaling differences in the correlated ground-state between left and right-handed chiral dye molecules introduced in the paper (red-dashed). The green line following $6.5 \cdot 10^{-8} \sqrt{N}$ serves as guide to the eye. It is apparent that the large-N limit, which is equivalent with increasing the fundamental coupling strength, is dominated by a $\sqrt{N}$ behavior that is consistent with the available literature. All parameters consistent with Fig.~3.}
    \label{fig:hopfield_linear}
\end{figure}

The Hopfield model has been shown to provide excellent results for vibrational strong coupling \cite{george2016} and intersubband transitions \cite{todorov2012}. It is not obvious that the same qualitative accuracy can be expected for the electronic subspace in atomic/molecular structures.
However, recent work by Riso et al.~\cite{riso2022strong} utilized an adjusted version of QED Coupled-cluster to estimate the discriminating strength of chiral fields on single and few molecules. They predict a $\sqrt{N}$ behavior of the discriminating strength in the correlated ground-state (Fig.~5 of Ref.~\cite{riso2022strong}) that is consistent with our observations shown in Fig.~\ref{fig:hopfield_linear} if the coupling is deep within the ultra-strong coupling domain.

\begin{figure}[h]
    \centering
\includegraphics[width=0.5\textwidth]{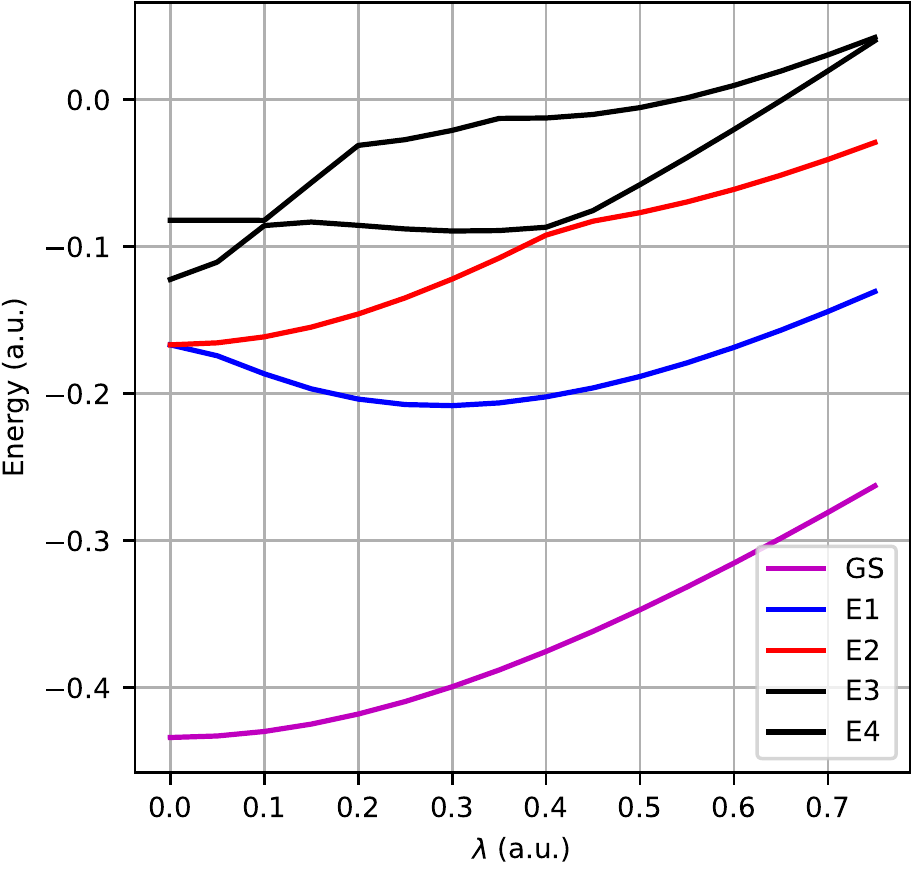}
    \caption{Eigenvalues of two-dimensional soft-Coulomb hydrogen coupled in electric dipole approximation to a single cavity mode in resonance with the first matter-excitation. The exact solution uses a grid representation with 151 x 151 grid-points for hydrogen and 30 Fock-states for the cavity mode. We show the first 5 correlated eigenstates. In total, this amounts to a Hilbert space with 684030 states. Our simplified Hopfield model provides a similar trend (as long as the magnetic components do not dominate the coupling) and is therefore qualitatively consistent with exact solutions. The exact solution will naturally produce (avoided) crossings between higher excited states that are not included in our model but that do not influence the drawn conclusions.}
    \label{fig:hydrogen2d}
\end{figure}

The overall trend of the polaritonic eigenvalues predicted by the Hopfield model is consistent with exact results for hydrogen as illustrated in Fig.~\ref{fig:hydrogen2d}.

\clearpage

\bibliography{Chirality}